%% file: Arxiv-template.tex
\documentclass[12pt]{article}
\usepackage[utf8]{inputenc}
\usepackage{pbox}

\usepackage{tikz}
\usetikzlibrary{shapes}
\usepackage{pgfplots}
\pgfplotsset{width=10cm,compat=1.9}

\usepgfplotslibrary{external}
\usepackage{tikz}
\usetikzlibrary{patterns,decorations.pathreplacing,calc,fadings,calc,shapes.callouts,shapes.arrows,backgrounds,spy,shadows,decorations.markings}
\immediate\write18{mkdir -p tikz}

\usetikzlibrary{intersections}
\usetikzlibrary{positioning}
\usetikzlibrary{decorations.pathreplacing}
\usetikzlibrary{matrix}
\usetikzlibrary{calc}
\usetikzlibrary{arrows,arrows.meta}
\usetikzlibrary{arrows,shapes,trees,backgrounds,shadows}
\usetikzlibrary{decorations.pathreplacing}
\usetikzlibrary{decorations.pathmorphing} 
\usetikzlibrary{backgrounds,snakes}
\usetikzlibrary{decorations.markings}
\usetikzlibrary{positioning}
\usetikzlibrary{plotmarks}
\usetikzlibrary{trees}
\usetikzlibrary{patterns}
\usetikzlibrary{spy}
\usetikzlibrary{calc,matrix}

\usepackage{scalerel}

\usepackage{graphicx} 
\usepackage[margin=1in]{geometry}
\usepackage{amsmath}
\usepackage{caption}
\usepackage{hyperref}
\usepackage{subcaption}
\usepackage{xcolor}
\usepackage{amssymb}
\usepackage[numbers]{natbib}
\usepackage{multirow}
\usepackage{soul}

\usepackage{amsmath, amsfonts, dsfont}
\usepackage{hyperref}
\usepackage{graphicx}
\usepackage{placeins}
\usepackage[ruled]{algorithm2e}

\newtheorem{theorem}{Theorem}
\newtheorem{lemma}{Lemma}

\newtheorem{assumption}{Assumption}

\newtheorem{definition}{Definition}

\newcommand{\argmin}{\mathrm{argmin}}

\input{notation}

\usepackage{xcolor}

\begin{document}




\title{Designing High-Occupancy Toll Lanes: A Game-Theoretic Analysis}


\author{%
    Zhanhao Zhang\\
    \footnotesize{Operations Research and Information Engineering, Cornell University, zz564@cornell.edu}\and
    Ruifan Yang\\
    \footnotesize{Operations Research and Information Engineering, Cornell University, ry298@cornell.edu}\and
    Manxi Wu\\
    \footnotesize{Department of Civil and Environmental Engineering, University of California, Berkeley, manxiwu@berkeley.edu}
}

\date{}
\maketitle

\begin{abstract}
    \input{abstract}
\end{abstract}



%


\input{intro}

\section{The basic model}
\input{model} \label{sec:model}

\section{Equilibrium characterization and comparative statics} \label{sub:one_segment}
\subsection{Equilibrium characterization} \label{sec:equilibrium}
\input{equilibrium}

\subsection{Comparative statics} \label{subsec:comparative_statics}
\input{comparative_statics}

\section{Extensions to multiple segments and occupancy levels} \label{sec:multiseg}
\input{multi_segments}

\section{Design HOT on California I-880} 
\label{sec:numerical}
\input{numerical}

\section{Concluding remarks}
\input{conclusion}




%
%
%


\bibliographystyle{unsrtnat} 
\bibliography{references} 

\clearpage

\appendix
\input{appendix}

\end{document}

%% file: notation.tex
\newcommand{\ubeta}{\overline{\beta}}
\newcommand{\ugamma}{\overline{\gamma}}

\newcommand{\poolmax}{\sigma^{\dagger}_{\mathrm{pool}}}

\newcommand{\toll}{\tau}

\newcommand{\actionset}{A}
\newcommand{\poolsize}{A}
\newcommand{\occreq}{A}
\newcommand{\actiont}{\mathrm{toll}}
\newcommand{\actionp}{\mathrm{pool}}
\newcommand{\actiono}{\mathrm{o}}

\newcommand{\vot}{\beta}
\newcommand{\disu}{\gamma}

\newcommand{\votmax}{\bar{\beta}}
\newcommand{\disumax}{\bar{\gamma}}

\newcommand{\pdf}{f}

\newcommand{\costh}[1]{\ell_{\mathrm{h}#1}}
\newcommand{\costo}[1]{\ell_{\mathrm{o}#1}}

\newcommand{\costhath}[1]{\hat{\ell}_{\mathrm{h}#1}}
\newcommand{\costhato}[1]{\hat{\ell}_{\mathrm{o}#1}}
\newcommand{\flowh}{x_{\mathrm{h}}}
\newcommand{\flowo}{x_{\mathrm{o}}}

\newcommand{\costa}{C_{\mathrm{a}}}

\newcommand{\flowheq}{\flowh^*}
\newcommand{\flowoeq}{\flowo^*}

\newcommand{\votset}{\mathrm{B}}
\newcommand{\disuset}{\Gamma}

\newcommand{\strategy}{s}
\newcommand{\region}[1]{R_{#1}}
\newcommand{\action}{a}

\newcommand{\strategydis}[1]{\sigma_{#1}}

\newcommand{\strategyeq}{s^*}
\newcommand{\strategydiseq}[1]{\sigma^*_{#1}}

\newcommand{\costdiff}{\ell_{\delta}}
\newcommand{\lthreshold}{\ell^{\dagger}_{\delta}}

\newcommand{\BR}{\mathrm{BR}}
\newcommand{\bestregion}[1]{\Lambda_{#1}}

\newcommand{\objcongestion}{\mathrm{C}}
\newcommand{\objemission}{\mathrm{E}}
\newcommand{\objutility}{\mathrm{U}}
\newcommand{\objrevenue}{\mathrm{R}}

\newcommand{\segnum}{e}
\newcommand{\segtotal}{E}

\newcommand{\demand}[2]{D^{#1}_{#2}}
\newcommand{\demandgrid}[2]{d^{#1}_{#2}}
\newcommand{\multisegflow}{x}
\newcommand{\multisegflowo}[1]{x_{\mathrm{o}#1}}
\newcommand{\multisegflowhot}[1]{x_{\mathrm{h}#1}}

\newcommand{\multisegflowohat}[1]{\hat{x}_{\mathrm{o},#1}}
\newcommand{\multisegflowhothat}[1]{\hat{x}_{\mathrm{h},#1}}

\newcommand{\multisegstrategyeq}[1]{s^{*#1}}
\newcommand{\multisegstrategydiseq}[2]{\sigma^{*#2}_{#1}}
\newcommand{\multisegfloweq}[2]{x^{*#2}_{#1}}
\newcommand{\multisegflowoeq}[2]{x_{\mathrm{o},#1}^{*#2}}
\newcommand{\multisegflowhoteq}[2]{x_{\mathrm{h},#1}^{*#2}}
\newcommand{\multisegflowoccu}[2]{\bar{x}^{#1}_{#2}}

\newcommand{\occumax}{M}
\newcommand{\occu}{m}
\newcommand{\occuratio}[2]{r_{#1}^{#2}}

\newcommand{\hotcap}{\rho}
\newcommand{\actionoccu}{a_{\mathrm{occu}}}
\newcommand{\actionseg}[1]{a_{#1}}

\renewcommand{\time}{t}
\newcommand{\Horizon}{T}
\renewcommand{\day}{n}
\newcommand{\Day}{N}

%% file: abstract.tex
In this article, we study the optimal design of High Occupancy Toll (HOT) lanes. The traffic authority determines the road capacity allocation between HOT lanes and ordinary lanes, as well as the toll price charged for travelers using HOT lanes who do not meet the high-occupancy eligibility criteria. We develop a game-theoretic model to analyze the decisions of travelers with heterogeneous preference parameters in values of time and carpool disutilities. These travelers choose between paying or forming carpools to use the HOT lanes, or taking the ordinary lanes. Travelers' welfare depends on the congestion cost of the lane they use, the toll payment, and the carpool disutilities. For highways with a single entrance and exit node, we provide a complete characterization of equilibrium strategies and a comparative statics analysis of how the equilibrium vehicle flow and travel time change with HOT capacity and toll price. We then extend the single segment model to highways with multiple entrance and exit nodes. We extend the equilibrium concept and propose various design objectives considering traffic congestion, toll revenue, and social welfare. Using the data collected from the HOT lane of the California Interstate Highway 880 (I-880), we formulate a convex program to estimate the travel demand and approximate the distribution of travelers' preference parameters. We then compute the optimal toll design of five segments for I-880 for achieve each one of the four objectives, and compare the optimal solution with the current toll pricing.

%% file: intro.tex
\section{Introduction}
High Occupancy Toll (HOT) lanes are traffic lanes or roadways that are open to vehicles satisfying a minimum occupancy requirement but also offer access to other vehicles with a toll price. In practice, HOT lanes have been implemented on several interstate highways in California, Texas, and Washington states. With the proper design of lane capacity and toll price, HOT lanes can effectively mitigate traffic congestion through incentivizing carpooling and transit use, while also generating revenue to support transportation infrastructure through toll collection.

The goal of our work is to study the optimal design of HOT lane systems and its impact on traffic congestion, social welfare, and revenue generated from toll collection. In our model, a traffic authority designs the HOT lane systems by choosing the road capacity of HOT lanes, and the toll price. Given the design of HOT, we develop a game-theoretic model to analyze the strategic decisions made by travelers who have the action set of paying or carpooling to use the HOT lane, or using the ordinary lane. Travelers are modeled as a population of nonatomic agents with a continuous distribution of value of time and carpool disutility. Both the HOT lanes and the ordinary lanes are congestible in that the travel time of each lane increases with the aggregate flow induced by agents' decisions. The outcomes of the system in terms of agent travel time cost and toll collection are jointly determined by agents' equilibrium strategies and the design by the traffic authority. 


In the first part of our work (Sec. \ref{sec:model}-\ref{sub:one_segment}), we consider highway segments with a single entrance and exit node. We provide a complete characterization of Wardrop equilibrium in this game. In particular, we identify two qualitatively distinct equilibrium regimes that depend on the traffic authority's design of lane capacity and toll price. In the first equilibrium regime, all agents who take the HOT lane form carpools and no one pays the toll due to the relatively high toll price. In the second equilibrium regime, a fraction of agents with high carpool disutilities and high value of times makes toll payment to take the HOT lanes, while the rest either forms carpools or takes the ordinary lanes. In both regimes, agents are split between taking the HOT lanes and the ordinary lanes. 

The equilibrium characterization provides the system designer with insights on how the equilibrium flows and travel time costs of both the HOT lanes and the ordinary lanes depend on the system parameters that include travel time cost functions, capacity allocation and toll price. Moreover, we present comparative static analysis on how the equilibrium flow and costs change with the fraction of capacity that is allocated to the HOT lanes and the toll price of HOT lanes. We find that if we increase the HOT capacity but holding the toll price fixed, the latency difference between ordinary lanes and HOT lanes increases. Moreover, more agents use the HOT lanes by paying the toll price or carpooling and fewer agents use ordinary lanes. On the other hand, increasing the toll price while holding the HOT capacity fixed will lead to an increase in the latency difference between ordinary lanes and HOT lanes, an increase in carpooling agents, and a decrease in toll-paying agents. The number of agents using ordinary lanes can change in either direction.

In the second part of our work (Sec. \ref{sec:multiseg}-\ref{sec:numerical}), we generalize our model to highways with multiple segments separated by entrance and exit nodes and carpool system with multiple occupancy levels. The toll price of each segment set by the system designer varies with the vehicles occupancy level. The population consists of agents with different entry and exit points. Agents choose their carpool occupancy levels before entering the highway, and can switch between ordinary lanes and HOT lanes for different road segments. We generalize our equilibrium concept to this model extension, and prove equilibrium existence. We also provide a generic sufficient condition under which equilibrium is unique.

We apply our model and equilibrium analysis in the numerical study using the data collected on the California I-880 HOT lane system, from Dixon Landing Road to Lewelling Boulevard. We calibrate the latency function using vehicle travel time and flow data provided by the Caltrans Performance Measurement System (PeMS). To compute the equilibrium strategy distribution, we need to estimate the demand for each entry and exit pair and the distribution of preference parameters among the population. To ensure tractability in estimating demand and preference distribution, we partition the preference parameter vector space into equally sized subsets, and estimate the demand of agents with preference in each subset building on the idea of inverse optimization. 
 That is, given the data on toll prices and driving time, we compute the equilibrium strategy profile of agents with all preference parameters and entrance and exit nodes. This allows us to map the demand estimate of each preference set to an induced vehicle flows on ordinary lanes and HOT lanes at equilibrium. We formulate a convex program to estimate the demand volumes as the one that minimizes the difference between the equilibrium vehicle flows and the observed flows on each lane and each segment. 

Next, we compute the equilibrium strategy profile for a set of discretized design parameters, including capacity allocation and toll prices. We consider four objective functions for the HOT design: (i) the total agent travel time, (ii) the total vehicle driving time, (iii) the total revenue measures the toll prices collected, and (iv) total cost of all agents taking into account their driving time, toll payments, and carpool disutilities. We select a time interval (5-6 pm) during the evening rush hour to compute the Pareto front for the design of HOT lanes under various toll prices and HOT capacities. This analysis illustrates the system authority's trade-off between reducing traffic congestion, enhancing social welfare, and maximizing total toll revenue. 
Charging a high toll price on road segments with higher demand is effective in incentivizing agents to carpool and reduce both agent travel time and vehicle driving time. On the other hand, for revenue maximization, a lower toll price is optimal to increase the fraction of toll-paying agents.

Moreover, we compute the optimal toll design under the current HOT capacity across all operating hours of a workday. Since demand volume is lower during the morning hours but higher in the afternoon, setting a high toll price in the afternoon is more effective for reducing agent travel time, vehicle driving time, and costs. However, for revenue maximization, it is optimal to set a low toll price throughout the day. By adjusting the current toll price to the optimal toll prices for each of these four objectives, we can achieve significant improvements in the corresponding objective.

\emph{Related Literature}. Our model and analysis build on the rich literature of congestion games that includes the equilibrium analysis of routing strategies made by atomic agents \cite{monderer1996potential, rosenthal1973class} and nonatomic agents \cite{sandholm2001potential} in networks, and the analysis on the price of anarchy \cite{roughgarden2004bounding, paccagnan2019incentivizing, correa2004selfish, roughgarden2005selfish}. Most of the classical results in congestion games have focused on the settings where all agents have homogeneous preferences. The papers \cite{milchtaich1996congestion, mavronicolas2007congestion} extended these results to study the equilibrium existence and efficiency with player-specific costs. Previous literature has also examined the optimal design of tolling mechanisms that minimize the social cost of nonatomic agents with homogeneous preferences \cite{christodoulou2005price, roughgarden2010algorithmic, roughgarden2002bad}. The optimal toll design with heterogeneous values of time has also been studied. For example, \cite{chen2004solving, liu2017credit, chen2015solving, liu2012welfare, ramos2020toll, hortelano2019optimal, karakostas2004edge, chen2015optimal, van2024self} consider a finite number of agent classes, where the value of time is the same among agents within the same class but differs across different classes. Furthermore, \cite{nie2010existence, yang2005mathematical, jiang2013toll, lu2006variable, fleischer2004tolls, coletolls} study the setting where the value of time of agents follow continuous distribution. These work do not incorporate carpooling into their models. 


There are extensive works modeling travelers' decisions regarding HOT/HOV lanes. The first stream of work examined the static user equilibrium of travel mode and/or route choices on a single-segment, multi-lane road. \cite{YANG} studied the user equilibrium with and without HOV lanes and discussed the optimal congestion pricing. 
\cite{Zang2018,Chu2012,Hughes2019} extended the HOV system with various schemes and settings. These works focus on HOV systems, which do not allow single occupancy vehicles to use the express lane. Moreover, they assume that all commuters have a homogeneous value of time and carpool disutility, except for \cite{Hughes2019}, which considered commuters with heterogeneous carpool disutilities but homogeneous value of time.
Considering the HOT systems that allow single occupancy vehicles to drive on the express lanes by paying a toll, \cite{jang2014bi, lou2011optimal, tan2018hybrid} discussed the pricing schemes where only single occupancy vehicles make decisions between HOT lanes and ordinary lanes while the number of high occupancy vehicles remains constant.   
\cite{Konishi2010}, allowing travelers to choose both their travel modes and routes simultaneously, compared the road efficiency for HOT and HOV systems, and explored the optimal congestion pricing scheme for both ordinary lanes and HOV lanes on a multi-lane highway. They also considered commuters with heterogeneous carpool disutilities but a homogeneous value of time. \cite{Yuan2024} studied the impact of ride-sourcing vehicles on both HOV and HOT systems, considering commuters with heterogeneous values of time but homogeneous carpool disutility. Additionally, \cite{Xu2015a, Xu2015b,  Yan2019b, Chong2022, Di2017, Di2018, Li2020, Li2021} extended these models to consider ridesharing user equilibrium (RUE) with HOT lanes. 
None of the above works considers commuters with both heterogeneous values of time and carpool disutilities, and they only considered homogeneous carpool occupancy levels.

More broadly, the second stream of works studied the impact of carpooling in a dynamic setting that extends the bottleneck model by \cite{Vickrey1969}, and examined departure time equilibrium and/or travel mode choice equilibrium with HOT/HOV lanes \citep{laval2015real}. 
 \cite{wu2011pareto, QianZhang} studied the HOV/HOT system with three traveling modes: transit, driving alone, and carpooling. They derived the departure time equilibrium for each traveling mode and how different factors affect the mode shares and network performance.
 \cite{Xiao2021b, Xiao2016, Wang2020, Liu2017, Wang2019, Zhong2020} extended the same problem to include parking space constraints or ride-sharing compensation.
However, all the above works considered only homogeneous commuters. 
\cite{Yu2019a} studied departure time equilibrium with heterogeneous users in the preference for cost of carpooling, values of time, and values of schedule delay. However, they modeled travelers' mode choice and route choice separately: although travelers achieve departure time equilibrium within each mode, the shares among the modes are determined by a nested logit model. Also, they only considered a single-lane scenario, which does not incorporate separate HOV/HOT lanes. Moreover, all the dynamic works above either assume a constant carpool occupancy level (\cite{Xiao2021a}, \cite{Xiao2021b}, \cite{QianZhang}, \cite{Wang2019}, \cite{Yu2019}), or carpool occupancy level that is a continuous variable as in \cite{Zhong2020}. 
The paper \cite{Wei2022} extended the temporal capacity allocation scheme with discrete heterogeneous occupancy level and heterogeneous carpool inconvenience costs, but considered homogeneous value of time.

Our work contributes to modeling and analysis of HOT lane systems in static settings. In particular, our model incorporates agents with heterogeneous values of time and heterogeneous carpool disutilities, where both parameters are continuously distributed. This dual-dimensional heterogeneity is essential for optimal HOT lane design since choices between ordinary and HOT lanes, and whether to pay or carpool, depend on both parameters. We proved equilibrium existence, uniqueness, and fully characterize the equilibrium structure with general latency functions and continuous preference distributions. Our results identify two distinct equilibrium regimes and lead to comparative statics that provide insights into how toll prices and HOT lane capacity effect lane usage and carpooling ratios. Furthermore, we extend our basic model to multi-segment settings with multiple occupancy levels and differentiated pricing. We prove equilibrium existence and uniqueness. These results support practical HOT lane design with multiple segments and tiered toll prices.

Moreover, our case study of the California I-880 highway contributes to the literature of numerical/empirical analysis on HOT design and carpooling. \cite{ekstrom2014optimal} investigated the impact of toll design on social welfare through a case study of Stockholm. \cite{fan2015optimal} and \cite{fan2017social} demonstrated the improvement of revenue and social welfare in the Sioux Falls network by adopting optimal tolling mechanisms. \cite{michalaka2013simulating, he2017optimal} used simulation to compute optimal toll prices across multiple objectives that incorporate time-of-day pricing, drivers' lane choice behaviors in the presence of tolls, and different toll structures across various road segments of HOT lane facilities. \cite{Cohen2022} and \cite{Cohen2023} designed field experiments and found a positive impact of HOV lanes on both carpooling intent and adoption. 
Additionally, \cite{Finkleman2011} surveyed 250+ drivers, and find relationships between willingness to pay and the improvement in travel speeds in HOT lanes, the length of the trip, and the urgency of on-time arrival. 

Our numerical study contributes to the above literature by bridging the game theory analysis of HOT lane system with data collected on the HOT project along California I 880 highway. 
In particular, we inversely estimated the distribution of travelers' preferences using the traffic sensor data and data requested from Caltrans on the aggregate lane choice and carpool ratios. We designed optimal tolls for four proposed objectives building on the equilibrium analysis and preference distribution estimates. Our results provide the optimal hourly toll pricing with the objectives of improving traffic congestion mitigation and toll revenue under different HOT capacities, and demonstrate the trade-off between different objectives.

%% file: model.tex
Consider a highway segment consisting of \emph{ordinary} and \emph{high occupancy toll} (HOT) lanes. An ordinary lane is toll-free and open to all vehicles. A high occupancy toll lane is accessible to vehicles that either pay the toll price $\toll \in \mathbb{R}_{\geq 0}$ or meet the minimum occupancy requirement with passenger size that is an integer $\occreq \geq 2$. A central planner (e.g. transportation authority) determines the toll price $\toll$, the minimum occupancy requirement $\occreq$, and the allocation of road capacity between HOT lanes and ordinary lanes. In particular, we denote the fraction of capacity allocated to HOT lanes as $\rho \in [0,1]$, and the remaining $(1 - \rho)$-fraction of capacity is allocated to the ordinary lanes. The capacity allocation affects the travel time cost (i.e. latency function) of the two types of lanes. We denote the latency function of the HOT lanes as $\costh{}(\flowh, \rho)$, and the latency function of the ordinary lanes as $\costo{}(\flowo, 1 - \rho)$, where $\flowh$ (resp. $\flowo$) is the flow of vehicles using the HOT lanes (resp. the ordinary lanes). We assume that the latency functions satisfy the following assumption: 
\begin{assumption}\label{as:cost}\hfill
\begin{enumerate}
\item[(a)] The latency function $\costo{}(\flowo, 1 - \rho)$ (resp. $\costh{}(\flowh, \rho)$) is increasing in the flow $\flowo$ (resp. $\flowh$), and increasing (resp. decreasing) in the capacity ratio $\rho$. 
\item[(b)] $\costo{}(0, 1 - \rho) = \costh{}(0, \rho)$ for any $\rho \in [0, 1]$.
\end{enumerate}
\end{assumption}
Assumption \ref{as:cost}(a) indicates that both lanes are congestible in that the latency increases as the flow increases.\footnote{This assumption is supported by queueing based models \cite{lu2018probabilistic,lu2022analytical} and empirical validations \cite{national1950highway}.} Additionally, the latency decreases in one type of lanes as the allocated capacity of that lane increases. Assumption \ref{as:cost}(b) implies that the free flow travel time, defined as the latency when the flow is zero, is the same for both types of lanes. This is a reasonable assumption since the free flow travel time is determined by the length of the highway and the speed limit.


 
We model travelers as non-atomic agents with a total demand of $D>0$. The action set of each agent is $\actionset = \{\actiont, \actionp, \actiono\}$, where $\actiont$ (resp. $\actionp$) is the action of taking the HOT lanes by paying the toll price (resp. by meeting occupancy requirement), and $\actiono$ is to take the ordinary lane. Agents have heterogeneous preferences about the travel time cost (relative to the monetary payment) as well as the disutility of forming carpool groups. We model the heterogeneous preferences of agents by the parameter of value of time (i.e. the amount of monetary cost that is equivalent to one unit time cost), denoted as $\vot \in \votset = [0, \votmax]$, and the carpool disutility, denoted as $\disu \in \disuset = [0, \disumax]$. The distribution of agents' preference parameters $(\vot, \disu)$ is represented by the probability density function $\pdf: \votset \times \disuset \to \mathbb{R}$ such that $\pdf(\vot, \disu)>0$ for all $(\vot, \disu)$ and $\int_{\votset\times \disuset} \pdf(\vot, \disu) d \vot d \disu=1$. 

We define the strategy of an agent as a mapping from their preference parameters $(\vot, \disu)$ to a pure strategy in action set $\actionset$, denoted as $\strategy: \votset \times \disuset \to \actionset$. The set of agents who choose each action $\action$, denoted by $\region{\action}$, is given by: {\small
\begin{align*}
\region{\action} = \left\{\votset \times \disuset| \strategy(\vot, \disu) =\action\right\}, \quad \forall \action \in \actionset.
\end{align*}
}
We represent the strategy distribution of the agent population as $\strategydis{} = (\strategydis{\action})_{\action \in \actionset}$, where {\small
\begin{align}\label{eq:strategydis}
\strategydis{\action} = \frac{1}{D}\int_{\region{\action}} \pdf(\vot, \disu) d\vot d\disu
\end{align}
}
is the fraction of agents who choose each action $\action \in \actionset$, and $\strategydis{\actiont}+ \strategydis{\actionp}+ \strategydis{\actiono}=1$. Here, both $\region{\action}$ and $\strategydis{\action}$ for each $\action$ depend on $\strategy$. We drop the dependence from the notation for simplicity. The flow on each type of lanes induced by $\strategydis{}$ is as follows: {\small
\begin{align}\label{eq:flow}
    \flowh = \left(\strategydis{\actiont} + \frac{\strategydis{\actionp}}{\occreq}\right) D, \quad \flowo = \strategydis{\actiono} D.
\end{align}
}
The cost of each agent with preference parameters $(\vot, \disu)$ for choosing actions $\actiont, \actionp, \actiono$ is given by: {\small
\begin{subequations}\label{eq:cost_functions}
    \begin{align}
    C_{\actiont}(\strategydis{}, \vot, \disu) &= \vot \cdot \costh{}\left(\flowh, \rho\right) + \toll, \label{subeq:cost_toll}\\
    C_{\actionp}(\strategydis{}, \vot, \disu) &= \vot \cdot \costh{}\left(\flowh, \rho\right) + \disu,\\
    C_{\actiono}(\strategydis{}, \vot, \disu) &= \vot \cdot \costo{}\left(\flowo, 1-\rho\right),
\end{align}
\end{subequations}
}
where $\vot \cdot \costh{}\left(\flowh, \rho\right)$ (resp. $\vot \cdot \costo{}\left(\flowo, 1-\rho\right)$) represents the cost of enduring the latency on the HOT lanes (resp. the ordinary lanes), and the cost of toll payment or the carpool disutility is added for action $\actiont$ and $\actionp$, respectively. 

A strategy profile $\strategyeq$ is a Wardrop equilibrium if no agent has incentive to deviate: 

\begin{definition}\label{def:eq} A strategy profile $\strategyeq: \votset \times \disuset \to \actionset$ is a Wardrop equilibrium if 
{\small
\begin{align*}
&\strategyeq(\vot, \disu) = \action,   \\
&\Rightarrow \quad C_{\action}(\strategydiseq{}, \vot, \disu) = \argmin_{\action' \in \actionset} C_{\action'}(\strategydiseq{}, \vot, \disu),\\ 
\quad& \forall (\vot, \disu) \in \votset \times \disuset,
\end{align*}
}
and $\strategydiseq{}$ is the associated equilibrium strategy distribution given by \eqref{eq:strategydis}. 
\end{definition}
That is, the action chosen by an agent with parameters $(\vot, \disu)$ in equilibrium minimizes their own cost compared to choosing the other two actions. Given equilibrium strategy distribution $\strategydiseq{}$, we denote the induced equilibrium flow on the HOT lanes and the ordinary lanes by $\flowheq$ and $\flowoeq$, respectively.

%% file: equilibrium.tex
In this section, we characterize the Wardrop equilibrium of the game. For ease of exposition, we define $\costdiff(\strategydis{}, \rho)$ as the difference of the latency between the ordinary lanes and the HOT lanes given the strategy distribution $\strategydis{}$ and the capacity allocation $\rho$: 
{\small
\begin{align}\label{eq:cost_difference}
    \costdiff(\strategydis{}, \rho):= \costo{}(\flowo, 1 - \rho) - \costh{}(\flowh, \rho),
\end{align}
}
where $\flowo$ and $\flowh$ are derived from $\strategydis{}$ as in \eqref{eq:flow}. 

We first show that the latency difference between the ordinary lanes and the HOT lanes is always non-negative. Furthermore, when the toll price is strictly positive, there will always be some agents taking the ordinary lane or carpooling.

\begin{lemma}\label{lem:possibleOutcomes}
If $\toll > 0$, then $\costdiff(\strategydiseq{}, \rho) > 0$, $\strategydiseq{\actionp} > 0$, and $\strategydiseq{\actiono} > 0$. 
\end{lemma}

We next characterize the best response strategies of agents for a given strategy distribution $\strategydis{}$. In particular, given $\strategydis{}$, the best response of an agent with parameter $(\vot, \disu)$ is the action that minimizes the associated cost as in \eqref{eq:cost_functions}. We denote the best response as $\BR(\strategydis{}, \vot, \disu) \in \actionset$. Then, we can separate the parameter set $\votset \times \disuset$ into three regions $\left(\bestregion{\action}(\strategydis{})\right)_{\action \in \actionset}$, where $\bestregion{\action}(\strategydis{}) := \{ \votset \times \disuset | \BR(\strategydis{}, \vot, \disu)  = \action\}$. The following lemma characterizes the three regions with respect to the latency difference $\costdiff(\strategydis{}, \rho)$ induced by $\strategydis{}$ and the toll price $\toll$:\footnote{We do not consider the boundary cases where the inequalities in Lemma \ref{lem:BestResponse} hold with equality. Agents with preference parameters on the boundaries of each region are indifferent between two or even all three actions. Their tie-breaking rule does not affect the equilibrium analysis, as agents are nonatomic, and the demand from agents with boundary preference parameters is effectively zero.}


\begin{lemma}\label{lem:BestResponse}
Given $\strategydis{}$, 
{\small
\begin{align*}
    \bestregion{\actiont}(\strategydis{}) &= \{\votset \times \disuset| \vot \costdiff(\strategydis{}, \rho) \geq \toll, \  \disu \geq \toll \}, \\
     \bestregion{\actionp}(\strategydis{})  &= \{\votset \times \disuset|  \vot  \costdiff(\strategydis{}, \rho)  \geq \disu, \  \disu \leq \toll \}, \\
     \bestregion{\actiono}(\strategydis{}) &= \{\votset \times \disuset| \vot  \costdiff(\strategydis{}, \rho)\leq \min\{\toll, \disu\} \}.
\end{align*}
}
\end{lemma}

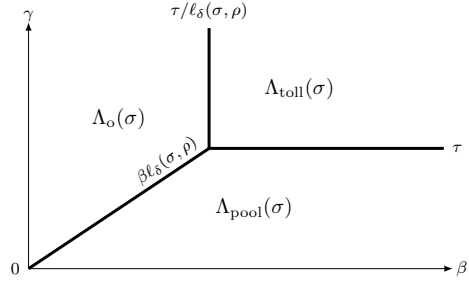
\begin{figure}[h]
    \centering
    \scalebox{0.8}{
        \begin{tikzpicture}[xscale=3, yscale=2]
            \draw[arrows={-latex}] (0,0) -- (2.35,0);
            \draw[arrows={-latex}] (0,0) -- (0,2.05);
            \node (00) at (2.4,0) {\scriptsize $\vot$};
            \node (01) at (0,2.1) {\scriptsize $\disu$};
            \node[left] (1) at (0,0) {\scriptsize 0};
            \node[right] (2) at (2.3,1) {\scriptsize $\toll$};
            \node[above] (3) at (1,2) {\scriptsize $\toll/\costdiff(\strategydis{}, \rho)$};
  
            \draw[line width=0.5mm, black] (0,0) -- (1,1);
            \draw[line width=0.5mm, black] (1,1) -- (1,2);
            \draw[line width=0.5mm, black] (1,1) -- (2.3,1);
            
            
            \node (4) at (1.5, 1.5) {\footnotesize $\bestregion{\actiont}(\strategydis{})$};
            \node (5) at (1.25, 0.5) {\footnotesize $\bestregion{\actionp}(\strategydis{})$};
            \node (6) at (0.5, 1.25) {\footnotesize $\bestregion{\actiono}(\strategydis{})$};
            \node[left] (7) at (1.0, 0.9) {\scriptsize \rotatebox{35}{$\beta \costdiff(\strategydis{}, \hotcap)$}};

        \end{tikzpicture}
    }
    \caption{Characterization of best response strategies}\label{fig:decisionBoundary}
 \end{figure}

We illustrate the three regions in Figure \ref{fig:decisionBoundary}. We note that $\bestregion{\actiont}(\strategydis{})$ includes agents with both high value of time $\vot$ and high carpool disutility $\disu$. Such agent prefers to take the HOT lanes via paying rather than taking the ordinary lanes due to their high value for time (i.e. the cost saving given $\costdiff(\strategydis{}, \rho)$ is no less than the toll price $\toll$), and also prefers to pay the toll price over carpooling due to their high carpool disutility (i.e. $\disu$ is no less than the toll price). Similarly, agents in $\bestregion{\actionp}(\strategydis{})$ have carpool disutility at most $\toll$, and thus prefer to carpool than paying the toll price. Their value of time $\vot$ is high relative to the carpool disutility $\disu$ so that the cost saving given $\costdiff(\strategydis{}, \rho)$ is no less than the carpool disutility, i.e. they prefer to take the HOT lanes by carpool rather than taking the ordinary lane. Finally, $\bestregion{\actiono}(\strategydis{})$ includes agents whose value of time is low relative to both their carpool disutility and toll price, and hence they prefer to take the ordinary lanes compared to taking the HOT lanes via carpool or toll payment.


We are now ready to present equilibrium characterization of this game. We define the following latency difference threshold value $\lthreshold$ that will be used for separating different equilibrium regimes: 
{\small
\begin{align}\label{eq:threshold}
    \lthreshold := \costdiff(\sigma^{\dagger}, \rho),
\end{align}
}
where $\sigma^{\dagger}$ is the threshold strategy distribution given by: 
{\small
\begin{subequations} \label{eq:poolmax}
    \begin{align}
        \sigma^{\dagger}_{\mathrm{toll}} =& 0,\\ 
        \poolmax =& \int_{0}^{\votmax} \int_0^{\min\{\toll,\ugamma\} \vot/\votmax} f(\vot, \disu) d\disu d\vot,\\ 
        \sigma^{\dagger}_{\mathrm{o}} =& 1- \poolmax.
    \end{align}
\end{subequations}
}
The following theorem shows that the game has a unique equilibrium that falls into either one of the two regimes depending on the game parameters. In each regime, the equilibrium strategy profile can be computed by solving a fixed point equation.



 

\begin{theorem} \label{thm:NE}
The game has a unique Wardrop equilibrium. 

\noindent  Regime A:  The toll price $\toll$ is relatively high, i.e. $\toll \geq \min\left\{\disumax, \votmax \lthreshold \right\}$. No agent takes HOT lanes by paying the toll, i.e. $\strategydiseq{\actiont} = 0$. Furthermore, 
 
 
\begin{enumerate}
	\item[(A-1)] If $\votmax \lthreshold \leq \ugamma$, 
 then $\strategydiseq{\actionp}$ is the unique solution that satisfies the following equation: {\small
	\begin{equation} 
		\strategydiseq{\actionp} = \int_{0}^{\votmax} \int_0^{\costdiff(0, \sigma^*_{\actionp}, 1 - \sigma^*_{\actionp}, \hotcap) \beta} f(\beta, \gamma) d\gamma d\beta, \label{eq:a1-fixed_point}
	\end{equation}		
        }
	and $\sigma^*_{\actiono} = 1 - \sigma^*_{\actionp}$.
	\item[(A-2)] If $\votmax \lthreshold > \ugamma$ 
 , then $\sigma^*_{\actionp}$ is the unique solution of the following equation: {\small
	\begin{equation} \label{eq:a2-fixed_point}
		1- \sigma^*_{\actionp} = \int_0^{\ugamma} \int_{0}^{\gamma/\costdiff(0, \sigma^*_{\actionp}, 1-\sigma^*_{\actionp}, \hotcap)} f(\beta, \gamma) d\beta d\gamma,
	\end{equation}		
        }
	 and $\sigma^*_{\actiono} = 1 -\sigma^*_{\actionp}$.
\end{enumerate}

\noindent \underline{Regime B}: The toll price $\toll$ is relatively low, i.e. $0 < \toll < \min\left\{\disumax, \votmax \lthreshold \right\}$. All three actions are taken by agents in equilibrium, and $\sigma^*$ is the unique solution that satisfies the following equations: {\small
\begin{subequations} \label{eq:b-fixed_point}
\begin{align}
	\sigma^*_{\actiont} =& \int_{\tau}^{\ugamma} \int_{\tau/\costdiff(\strategydiseq{}, \hotcap)}^{\ubeta} f(\beta, \gamma) d\beta d\gamma, \label{eq:b-fixed_point-toll}\\
	\sigma^*_{\actionp} =& \int_{0}^{\tau} \int_{\gamma/\costdiff(\strategydiseq{}, \hotcap)}^{\ubeta} f(\beta, \gamma) d\beta d\gamma, \label{eq:b-fixed_point-pool}\\
	\sigma^*_{\actiono} =& 1 - (\sigma^*_{\actiont} + \sigma^*_{\actionp}). \label{eq:b-fixed_point-o}
\end{align}
\end{subequations}
}


\end{theorem}

\medskip 

We provide the proof intuition of Theorem \ref{thm:NE} in this section. The complete proof is in Appendix \ref{apx:proof}. Our equilibrium characterization builds on the two lemmas \ref{lem:possibleOutcomes} -- \ref{lem:BestResponse} introduced before. In particular,
Lemma \ref{lem:possibleOutcomes} shows that in equilibrium both lanes are used, and either (A) all agents who take the HOT lanes choose to carpool, or (B) a positive fraction of agents who take the HOT lanes pay the toll $\tau$. Indeed, (A) and (B) are each associated with equilibrium regimes A and B, respectively. 

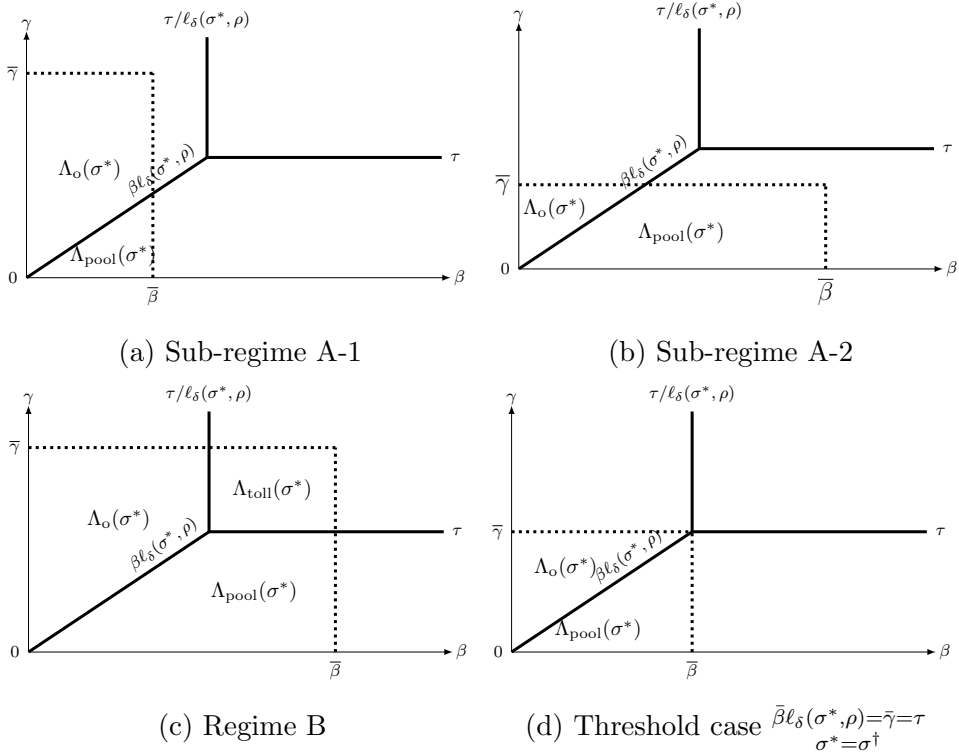
\begin{figure}[ht]
    \centering    
    \subfloat[\centering Sub-regime A-1]{\scalebox{0.8}{%
	\begin{tikzpicture}[xscale=3,yscale = 2]  
		\draw[arrows={-latex}] (0,0) -- (2.35,0);
		\draw[arrows={-latex}] (0,0) -- (0,2.05);
		\node (00) at (2.4,0) {\scriptsize $\beta$};
		\node (01) at (0,2.1) {\scriptsize $\gamma$};
		\node[left] (1) at (0,0) {\scriptsize $0$};
		\node[right] (2) at (2.3,1) {\scriptsize $\toll$};
		\node[above] (3) at (1,2) {\scriptsize $\toll/\costdiff(\strategydiseq{}, \hotcap)$};
            \node[left] (4) at (1.0, 0.9) {\scriptsize \rotatebox{35}{$\beta \costdiff(\strategydiseq{}, \hotcap)$}};
            \node[below] (7) at (0.7,0) {\scriptsize $\ubeta$};
            \node[left] (8) at (0,1.7) {\scriptsize $\ugamma$}; 
		\draw[line width=0.5 mm, black]  (0,0)  -- (1,1);
            \draw[line width=0.5 mm, black]  (1,1)  -- (1,2);
            \draw[line width=0.5 mm, black]  (1,1)  -- (2.3,1);
            \draw[line width=0.5 mm, black, dotted]  (0,1.7)  -- (0.7,1.7);
            \draw[line width=0.5 mm, black, dotted]  (0.7,0)  -- (0.7,1.7);
            \node (5) at (0.48, 0.18) {\footnotesize $\Lambda_{\mathrm{pool}}(\sigma^*)$};
            \node (6) at (0.35, 0.9) {\footnotesize $\Lambda_{\mathrm{o}}(\sigma^*)$};
	\end{tikzpicture}
} \label{subfig:regimes-A1}}
    \subfloat[\centering Sub-regime A-2]{\scalebox{0.8}{%
	\begin{tikzpicture}[xscale=3,yscale = 2]  
		\draw[arrows={-latex}] (0,0) -- (2.35,0);
		\draw[arrows={-latex}] (0,0) -- (0,2.05);
		\node (00) at (2.4,0) {\scriptsize $\beta$};
		\node (01) at (0,2.1) {\scriptsize $\gamma$};
		\node[left] (1) at (0,0) {\scriptsize $0$};
		\node[right] (2) at (2.3,1) {\scriptsize $\toll$};
		\node[above] (3) at (1,2) {\scriptsize $\toll/\costdiff(\strategydiseq{}, \hotcap)$};
            \node[left] (4) at (1.0, 0.9) {\scriptsize \rotatebox{35}{$\beta \costdiff(\strategydiseq{}, \hotcap)$}};
  
            \node[below] (7) at (1.7,0) { $\ubeta$};
            \node[left] (8) at (0,0.7) { $\ugamma$};
  
		\draw[line width=0.5 mm, black]  (0,0)  -- (1,1);
            \draw[line width=0.5 mm, black]  (1,1)  -- (1,2);
            \draw[line width=0.5 mm, black]  (1,1)  -- (2.3,1);
            \draw[line width=0.5 mm, black, dotted]  (0,0.7)  -- (1.7,0.7);
            \draw[line width=0.5 mm, black, dotted]  (1.7,0)  -- (1.7,0.7);
            \node (5) at (0.9, 0.3) {\footnotesize $\Lambda_{\mathrm{pool}}(\sigma^*)$};
            \node (6) at (0.2, 0.5) {\footnotesize $\Lambda_{\mathrm{o}}(\sigma^*)$};
	\end{tikzpicture}  
}\label{subfig:regimes-A2}}\\
    \subfloat[\centering Regime B]{\scalebox{0.8}{%
	\begin{tikzpicture}[xscale=3,yscale = 2]  
		\draw[arrows={-latex}] (0,0) -- (2.35,0);
		\draw[arrows={-latex}] (0,0) -- (0,2.05);
		\node (00) at (2.4,0) {\scriptsize $\beta$};
		\node (01) at (0,2.1) {\scriptsize $\gamma$};
		\node[left] (1) at (0,0) {\scriptsize $0$};
		\node[right] (2) at (2.3,1) {\scriptsize $\tau$};
		\node[above] (3) at (1,2) {\scriptsize $\tau/\costdiff(\strategydiseq{}, \hotcap)$};
            \node[left] (4) at (1.0, 0.9) {\scriptsize \rotatebox{35}{$\beta \costdiff(\strategydiseq{}, \hotcap)$}};
            \node[below] (7) at (1.7,0) {\scriptsize $\ubeta$};
            \node[left] (8) at (0,1.7) {\scriptsize $\ugamma$};   
		\draw[line width=0.5 mm, black]  (0,0)  -- (1,1);
            \draw[line width=0.5 mm, black]  (1,1)  -- (1,2);
            \draw[line width=0.5 mm, black]  (1,1)  -- (2.3,1);
            \draw[line width=0.5 mm, black, dotted]  (0,1.7)  -- (1.7,1.7);
            \draw[line width=0.5 mm, black, dotted]  (1.7,0)  -- (1.7,1.7);    
		\node (4) at (1.35, 1.35) {\footnotesize $\Lambda_{\mathrm{toll}}(\sigma^*)$};
            \node (5) at (1.25, 0.5) {\footnotesize $\Lambda_{\mathrm{pool}}(\sigma^*)$};
            \node (6) at (0.5, 1.1) {\footnotesize $\Lambda_{\mathrm{o}}(\sigma^*)$};
	\end{tikzpicture}
    }\label{subfig:regimes-B}}
    \subfloat[\centering Threshold case 
 $\substack{\bar{\beta} \ell_{\delta}(\sigma^*, \rho)=\bar{\gamma}=\tau \\ \sigma^* = \sigma^{\dagger}}$]{\scalebox{0.8}{%
    	\begin{tikzpicture}[xscale=3,yscale = 2]  
		\draw[arrows={-latex}] (0,0) -- (2.35,0);
		\draw[arrows={-latex}] (0,0) -- (0,2.05);
		\node (00) at (2.4,0) {\scriptsize $\beta$};
		\node (01) at (0,2.1) {\scriptsize $\gamma$};
		\node[left] (1) at (0,0) {\scriptsize $0$}; 
		\node[right] (2) at (2.3,1) {\scriptsize $\toll$};
		\node[above] (3) at (1,2) {\scriptsize $\toll/\costdiff(\strategydiseq{}, \hotcap)$};
            \node[left] (4) at (0.9, 0.8) {\scriptsize \rotatebox{35}{$\beta \costdiff(\strategydiseq{}, \hotcap)$}};
            \node[below] (7) at (1,0) {\scriptsize $\ubeta$};
            \node[left] (8) at (0,1) {\scriptsize $\ugamma$};
		\draw[line width=0.5 mm, black]  (0,0)  -- (1,1);
            \draw[line width=0.5 mm, black]  (1,1)  -- (1,2);
            \draw[line width=0.5 mm, black]  (1,1)  -- (2.3,1);
            \draw[line width=0.5 mm, black, dotted]  (0,1)  -- (1,1);
            \draw[line width=0.5 mm, black, dotted]  (1,0)  -- (1,1);
		\node (5) at (0.48, 0.18) {\footnotesize $\Lambda_{\mathrm{pool}}(\sigma^*)$};
            \node (6) at (0.30, 0.7) {\footnotesize $\Lambda_{\mathrm{o}}(\sigma^*)$};	
	\end{tikzpicture}
    }\label{subfig:regimes-lthres}}
    \caption{Equilibrium outcomes in each regime.}
    \label{fig:regimes}
\end{figure}

Furthermore, following Definition \ref{def:eq}, an equilibrium strategy distribution $\strategydiseq{}$ must satisfy {\small
\begin{equation} \label{eq: equilibrium condition}
    \strategydiseq{\action} = \iint_{\bestregion{\action}(\strategydiseq{})}  \pdf(\vot,\disu) d\vot d\disu,  \quad \action \in \actionset,
\end{equation}
}
where $\bestregion{\action}(\strategydiseq{})$ is the best response region characterized in Lemma \ref{lem:BestResponse}. 
In particular, using the best response characterization in Lemma \ref{lem:BestResponse}, the equilibrium distribution of carpool   $\strategydiseq{\actionp}$ can be written as: {\small \[
    \strategydiseq{\actionp} = \int_{0}^{\ubeta} \int_{0}^{\min\left\{\costdiff(\strategydiseq{}, \hotcap) \vot, \toll, \ugamma \right\}} \pdf(\vot, \disu) d\disu d\vot.
\]}
The two sub-regimes, (A-1) and (A-2), and regime B each corresponds to the scenario where one of the three elements, $\votmax \costdiff(\strategydiseq{}, \hotcap)$, $\toll$, or $\ugamma$, is the smallest. 
In particular, sub-regime A-1 (resp. A-2) corresponds to the case where $\votmax \costdiff(\strategydiseq{}, \hotcap)$ (resp. $\ugamma$) is the smallest among three elements. 
Thus, no agents use the HOT lane by paying the toll price since the toll price $\tau$ is either larger than the value of the time saved by taking the HOT lane $\votmax \costdiff(\strategydiseq{}, \hotcap)$ or larger than the maximum carpool disutility $\bar{\gamma}$. Figures \ref{subfig:regimes-A1} -- \ref{subfig:regimes-A2} illustrate that, in the equilibrium of sub-regimes A-1 and A-2, agents only choose to take the ordinary lane or carpool to take the HOT lane. This is because the preference parameter set does not intersect with the set corresponding to choosing to pay the toll to take the HOT lane as the best response strategy. 
In regime B, $\toll$ is the smallest element of the three (i.e. $0 < \tau < \min \{\bar{\gamma}, \bar{\beta}\ell_{\delta}(\sigma^*, \rho)\}$ as shown in Figure \ref{subfig:regimes-B}), and the strategy distributions for all three actions are positive in equilibrium. 


Finally, the threshold strategy distribution $\sigma^{\dagger}$ as in \eqref{eq:poolmax} and the threshold latency cost difference $\ell^{\dagger}_{\delta}$ as in \eqref{eq:threshold} are derived from the boundary case $\bar{\beta} \ell_{\delta}(\sigma^*, \rho) = \bar{\gamma} = \tau$, see Fig. \ref{subfig:regimes-lthres} for the illustration. 
In this threshold case, $\sigma^* = \sigma^{\dagger}$  and the latency difference $\ell_{\delta}(\sigma^*, \rho)= \ell^{\dagger}_{\delta}$. We can verify that this boundary case indeed marks the transition between different regimes. For example, in sub-regime A-1, 
{\small
\begin{align*}
    \strategydiseq{\actionp} =& \int_{0}^{\ubeta} \int_{0}^{\min\left\{\costdiff(\strategydiseq{}, \hotcap) \vot, \toll, \ugamma \right\}} \pdf(\vot, \disu) d\disu d\vot\\ 
    \stackrel{(a)}{=}& \int_{0}^{\ubeta} \int_{0}^{\costdiff(\strategydiseq{}, \hotcap) \vot} \pdf(\vot, \disu) d\disu d\vot\\ 
    \stackrel{(b)}{\leq}& \int_{0}^{\ubeta} \int_{0}^{\min\{\toll, \ugamma\} \vot / \votmax} \pdf(\vot, \disu) d\disu d\vot = \poolmax,
\end{align*}
}
and $\ell_{\delta}(\sigma^*, \rho) \leq \ell^{\dagger}_{\delta}$, where both (a) and (b) are due to the sub-regime A-1 conditions. Therefore, the sub-regime A-1 boundary characterization $\ugamma \geq \bar{\beta} \ell^{\dagger}_{\delta}$ guarantees that the sub-regime equilibrium condition $\ugamma \geq \bar{\beta} \ell_{\delta}(\sigma^*, \rho)$ holds when problem instance parameters are in A-1. Similarly, we can show that $\strategydiseq{\actionp} \geq \poolmax$ and $\ell_{\delta}(\sigma^*, \rho) \geq \ell^{\dagger}_{\delta}$ in sub-regime A-2 and regime B. As a result, the sub-regime A-2 (resp. regime B) characterization $ \bar{\gamma} \leq  \bar{\beta} \ell^{\dagger}_{\delta}$ (resp. $0< \tau < \min\{\bar{\gamma}, \bar{\beta}\ell^{\dagger}_{\delta}\}$) guarantees that the condition $ \bar{\gamma} \leq \bar{\beta} \ell^*_{\delta}(\sigma^*, \rho)$ (resp. $0< \tau < \min\{\bar{\gamma}, \bar{\beta}\ell_{\delta}(\sigma^*, \rho) \}$) is satisfied when problem instance parameters are in A-2 (resp. regime B). The detailed description of these conditions of each regime is in the theorem proof in Appendix \ref{apx:proof}.




\color{black}

%% file: comparative_statics.tex
We analyze the change of equilibrium strategy distribution with HOT capacity $\rho$ and toll price $\toll$. 



\begin{theorem}\label{theorem:comparative}
    The comparative statics of equilibrium strategy distribution $\strategydiseq{}$ with respect to $(\rho, \toll)$ are summarized in Table \ref{tab:comparative_statics}. 
    \begin{table}[htb]
        \centering
        \begin{tabular}{|c|c|c|}
            \hline
             & Fix $\toll$ increase $\hotcap$ & Fix $\hotcap$ increase $\toll$ \\
            \hline
            $\strategydiseq{\actiono}$ & Decreasing & Either Direction\\
            $\strategydiseq{\actiont}$ & Increasing & Non-Increasing\\
            $\strategydiseq{\actionp}$ & Increasing & Non-Decreasing\\
            $\costdiff(\strategydiseq{}, \hotcap)$ & Increasing & Non-Decreasing\\
            \hline
        \end{tabular}
        \caption{Comparative statics.}
        \label{tab:comparative_statics}
    \end{table}
\end{theorem}


Intuitively, if we fix the toll price $\tau$ and increase the HOT capacity $\rho$, some agents will switch from ordinary lanes to HOT lanes by either paying the toll price or carpooling. Therefore, $\strategydiseq{\actiont}$ increases, $\strategydiseq{\actionp}$ increases, and $\strategydiseq{\actiono}$ decreases. From analyzing the regime boundary change, we can also conclude that the latency difference between the ordinary lanes and the HOT lanes increases. 

On the other hand, if we fix the HOT capacity $\rho$ and increase the toll price $\toll$, some agents will deviate from paying the toll to carpooling or taking the ordinary lane. Hence, the latency in the ordinary lanes will increase and in the HOT lanes will decrease, which means the latency difference between the two lanes $\costdiff(\strategydiseq, \hotcap)$ will increase. Additionally, $\strategydiseq{\actiont}$ decreases and $\strategydiseq{\actionp}$ increases. As for the ordinary lanes, agents with high value of time and low carpool disutilities will switch to carpools as HOT lane becomes less congested. Depending on the preference distribution and the toll prices before and after the change, we can either have more agents switch from ordinary lanes to carpool or more agents switch from toll paying to ordinary lanes. Hence, $\strategydiseq{\actiono}$ can either increase or decrease. We remark that when toll price is already higher than the threshold where no agents pay the toll (regime A in Theorem \ref{thm:NE}), further increasing the toll price has no impact on the strategy distributions and therefore does not affect the latency difference.


%% file: multi_segments.tex
Consider a highway partitioned into multiple segments by separation nodes, which represent the locations where vehicles get on and off the highway. We number the segments sequentially from upstream to downstream as $\segnum \in [\segtotal]:=1, \dots, \segtotal$, where $\segtotal$ is the total number of road segments. Same as the basic model, the central planner divides the capacity of the highway into the HOT lane and the ordinary lane, and $\hotcap \in [0, 1]$ is the fraction of capacity allocated to the HOT lane. Agents can form carpools with different occupancy levels $\occu \in [\occumax] := 1, \dots, \occumax$, where $\occu= 1$ indicates that the agent does not carpool with others, and $\occumax$ is the maximum carpool size. For each segment $\segnum \in [\segtotal]$, a toll price $\toll_{\segnum, \occu} \geq 0$ is charged on each vehicle using the HOT lane with occupancy level $\occu$ on segment $\segnum$. Agents split the toll price evenly, i.e. a agent pays $\toll_{\segnum, \occu}/m$ when carpooling with $\occu-1$ other agents on the HOT lane of segment $\segnum$. The latency function on each road segment $\segnum$ is given by $\costo{,\segnum}\left(\multisegflowo{,\segnum}, 1 - \hotcap \right)$ for ordinary lanes and $\costh{,\segnum}\left(\multisegflowhot{,\segnum}, \hotcap \right)$ for HOT lanes, where $\multisegflowo{,\segnum}$ and $\multisegflowhot{,\segnum}$ are the vehicle flows on ordinary lanes and HOT lanes of road segment $\segnum$, respectively. Same as the basic model, we assume that the latency functions on all road segments satisfy Assumption \ref{as:cost}.

For each pair of $i \leq j \in [\segtotal]$, a population of agents with demand $\demand{ij}{}$ enters the highway from the entrance node (the beginning) of segment $i$ and leaves the highway from the exit node (the ending) of segment $j$. We refer the population who traverses segments $[i:j]:=i, \dots, j$ as the population $(i, j)$. Agents in each population $(i, j)$ decides their carpool size $m \in [M]$ and whether to take the HOT lane or the ordinary lane in each segment $\segnum \in [i:j]$. We denote an action of population $(i, j)$ as $\action = (\actionoccu, (\actionseg{\segnum})_{\segnum \in [i, j]})$, where $\actionoccu \in [M]$ is the occupancy level, and $\actionseg{\segnum} \in \{\mathrm{o}, \mathrm{h}\}$ is to take the ordinary lane or the HOT lane for each segment $\segnum \in [i, j]$. Thus, the action set of the population $(i, j)$ is $A^{ij} = [\occumax] \times \{\mathrm{o}, \mathrm{h}\}^{[i:j]}$. We note that agents must select a single carpool size for traversing all the segments but they can switch between the ordinary lanes and the HOT lanes at the separation nodes in between segments based on the HOT toll price and latency of the next segment. 

Analogous to the basic model, we represent the heterogeneous preference of agents of using value of time $\vot \in [0, \ubeta]$ and carpool disutilities $\disu := \left(\disu_{\occu} \right)_{\occu \in [\occumax]}$, where $\disu_{\occu} \in [0, \ugamma_{\occu}]$ denotes the disutilities for choosing occupancy level $\occu$. We set the disutility of single occupancy $\disu_1$ to $0$. The distribution of agents' preference parameters $(\vot, \disu)$ for population $(i, j)$ is represented by the probability density function $\pdf^{ij}: \votset \times \disuset \to \mathbb{R}$ such that $\pdf^{ij}(\vot, \disu)> 0$ for all $(\vot, \disu)$ and $\int_{\votset\times \disuset} \pdf^{ij}(\vot, \disu) d \vot d \disu=1$. 

We define the strategy of an agent in population $(i, j)$ as a mapping from their preference parameters $(\vot, \disu)$ to a pure strategy in action set $\actionset^{ij}$, denoted as $\strategy^{ij}: \votset \times \disuset \to \actionset^{ij}$. The set of agents of the population $(i, j)$ who choose each action $\action \in \actionset^{ij}$, denoted by $\region{\action}^{ij}$, is given by: 
{\small
\begin{align}\label{eq:sigmaij}
\region{\action}^{ij} = \left\{\votset \times \disuset| \strategy^{ij}(\vot, \disu) =\action\right\}, \quad \forall \action \in \actionset^{ij}.
\end{align}
}
We represent the strategy distribution of the population $(i, j)$ as $\strategydis{}^{ij} = (\strategydis{\action}^{ij})_{\action \in \actionset^{ij}}$, where {\small
\begin{align}\label{eq:multiseg-strategydis}
\strategydis{\action}^{ij} = \frac{1}{\demand{ij}{}}\int_{\region{\action}^{ij}} \pdf^{ij}(\vot, \disu) d\vot d\disu
\end{align} 
}
is the fraction of agents who choose each action $\action \in \actionset$, and $\sum_{\action \in \actionset^{ij}} \strategydis{\action}^{ij} = 1$. On each road segment $\segnum$, the vehicle flow on ordinary lanes $\multisegflowo{,\segnum}$ and on HOT lanes $\multisegflowhot{,\segnum}$ are induced by all agents of population $(i, j)$ that enters the highway on or before $\segnum$ and exits on or after $\segnum$, i.e. $i \leq \segnum \leq j$. In particular, {\small
\begin{subequations} \label{eq:multiseg-flow}
    \begin{align}
        \multisegflowo{,\segnum} =& \sum_{i = 1}^{\segnum}\sum_{j = \segnum}^{\segtotal} \demand{ij}{} \left(\sum_{\action \in \actionset^{ij}} \frac{1}{\actionoccu} \strategydis{a}^{ij} \mathds{1}_{\actionseg{\segnum} = \mathrm{o}} \right),\quad \forall \segnum \in [\segtotal],\\
        \multisegflowhot{,\segnum} =& \sum_{i = 1}^{\segnum}\sum_{j = \segnum}^{\segtotal} \demand{ij}{} \left(\sum_{\action \in \actionset^{ij}} \frac{1}{\actionoccu} \strategydis{a}^{ij} \mathds{1}_{\actionseg{\segnum} = \mathrm{h}} \right),\quad \forall \segnum \in [\segtotal].
    \end{align}
\end{subequations}
}

The cost of each agent of population $(i, j)$ with preference parameters $(\vot, \disu)$ for choosing action $\action \in \actionset^{ij}$ is given by
{\small
\begin{align} \label{eq:multiseg-cost}
    &\costa(\strategydis{}, \vot, \disu) = \disu_{\actionoccu}\nonumber\\
    +& \sum_{\segnum = i}^j \vot \cdot \costo{,\segnum}(\multisegflowo{,\segnum}, 1 - \hotcap) \mathds{1}_{\actionseg{\segnum} = \mathrm{o}} \notag \\
    +& \sum_{\segnum = i}^j \left(\vot \cdot \costh{,\segnum}(\multisegflowhot{,\segnum}, \hotcap) + \frac{1}{\actionoccu} \toll_{\segnum, \actionoccu} \right) \mathds{1}_{\actionseg{\segnum} = \mathrm{h}},
\end{align}
}
where $\vot \cdot \costo{,\segnum}(\multisegflowo{,\segnum}, 1 - \hotcap)$ (resp. $\vot \cdot \costh{,\segnum}(\multisegflowhot{,\segnum}, \hotcap)$) represents the cost of enduring the latency on ordinary lanes (resp. HOT lanes) on segment $\segnum$. Additionally, the carpool disutility $\disu_{\actionoccu}$ is added according to the associated occupancy level of the action, and toll payment $ \frac{1}{\actionoccu} \toll_{\segnum, \actionoccu} $ is added based on whether the action takes the HOT lane on each segment. Analogous to the basic model, we define the Wardrop equilibrium as:

\begin{definition}\label{def:multiseg-eq} A strategy profile $\multisegstrategyeq{}: \votset \times \disuset \to \actionset$ is a Wardrop equilibrium if
{\small
\begin{align*}
    &\multisegstrategyeq{ij}(\vot, \disu) = \action, \\
    &\Rightarrow \quad C_{\action}(\multisegstrategydiseq{}{ij}, \vot, \disu) = \argmin_{\action' \in \actionset^{ij}} C_{\action'}(\multisegstrategydiseq{}{ij}, \vot, \disu), \\
    &\forall (\vot, \disu) \in \votset \times \disuset,\ \forall i\leq j \in [\segtotal],
\end{align*}
}
and $\multisegstrategydiseq{}{ij}$ is equilibrium strategy distribution of population $(i, j)$ induced by $s^*$ given by \eqref{eq:sigmaij} -- \eqref{eq:multiseg-strategydis}. 
\end{definition}

We next provide conditions under which equilibrium is unique in the multi-segment setting. Before presenting this result, we note that an agent's equilibrium strategy $s^{*ij}(\beta, \gamma)$ depends on the segment latency of each lane, but only through the difference between them, not their individual values. We define $\delta = (\delta_e)_{e \in E}$ as the vector of latency difference, where $\delta_e$ is the latency of the ordinary lane exceeding that of the HOT lane in segment $e$. Given $\delta$ and the toll price vector $\tau$, the best response of agent in population $(i,j)$ with preference parameter $(\beta, \gamma)$ is uniquely determined. In particular, when an agent with preference parameter $(\beta, \gamma)$ chooses occupancy level $m$, they will select the ordinary lane (resp. HOT lane) of segment $e$ if $\toll_{\segnum, m}/m > \beta \delta_e$ (resp. $\toll_{\segnum, m}/m < \beta \delta_e$). The agent's best response occupancy level is $\argmin_{m \in [M]} \{\sum_{e \in [i, j]} \min \{ \toll_{\segnum, m}/m, \ \beta \delta_e \} + \gamma_m\}$. With slight abuse of notation, we denote $x(\delta)$ as the lane flow vector induced by all agents taking their best response to the cost difference vector $\delta$ according to \eqref{eq:multiseg-strategydis}-\eqref{eq:multiseg-flow}. We define $\Phi_e(\delta) = \ell_{o,e}(x_{o, e}(\delta)) - \ell_{h, e}(x_{h, e}(\delta))$ as the latency cost difference between the two lanes of segment $e$ induced by all agents' best response given $\delta$, and the vector function is $\Phi(\delta)= (\Phi_e(\delta))_{e \in E}$. 

We note that the set of fixed point solution of $\Phi(\delta) = \delta$ is the vector of latency cost difference in equilibrium $\delta^*$. This is because when the latency cost difference induced by agents' best response $\Phi(\delta^*)$ is consistent with the actual latency cost difference $\delta^*$, no agent has incentive to deviate. Moreover, since agents have unique best response for any $\delta$, there is a one-to-one correspondence between an equilibrium strategy profile $s^*$ and an equilibrium latency cost difference vector $\delta^*$. The following theorem shows that in the multi-segment model, equilibrium exists, and is unique when the function $\Phi(\cdot)$ satisfies certain condition.

\begin{theorem} \label{thm:multiseg-equi-uniqueness}
    Given any toll price vector $\tau$, Wardrop equilibrium $\multisegstrategyeq{}{}$ exists. Moreover, $\multisegstrategyeq{}{}$ is unique if the Jacobian matrix $\triangledown \Phi(\delta)$ does not have $1$ as its eigenvalue for any $\delta \in \prod_{e \in E} [\underline{\delta}_e, \bar{\delta}_e]$, where $\underline{\delta}_e= \ell_{o, e}(0) -\costh{,\segnum}\left(\sum_{i = 1}^{\segnum} \sum_{j = \segnum}^{\segtotal} D^{ij} \right)$ and $\bar{\delta}_e = \costo{,\segnum}\left(\sum_{i = 1}^{\segnum} \sum_{j = \segnum}^{\segtotal} D^{ij} \right) - \costh{,\segnum}(0)$. 
\end{theorem}
In Theorem \ref{thm:multiseg-equi-uniqueness}, the equilibrium existence result is established using the Kakutani's fixed point theorem, relying on the boundedness of the latency cost difference vector and the continuity of the function $\Phi(\cdot)$. The equilibrium uniqueness result follows from the one-to-one correspondence between equilibrium and fixed point solution of $\Phi(\delta)= \delta$, and the mean value theorem. We note that the sufficient condition holds generically, meaning that equilibrium is generically unique. Additionally, in the single-segment setting, we can verify that $\Phi(\delta)$ is monotone in $\delta$ and that $\nabla \Phi(\delta)$ does not have an eigenvalue of 1. As demonstrated in Theorem \ref{thm:NE}, equilibrium is indeed unique in the single segment setting. The complete proof of Theorem \ref{thm:multiseg-equi-uniqueness} can be found in Appendix \ref{apx:proof}.

%% file: numerical.tex
In this section, we redesign the high-occupancy toll (HOT) lane on California's I-880. Using data collected from the HOT operations on I-880 in 2021, we calibrate a multi-segment model. The primary challenge in this calibration is estimating the distribution of agents' preferences from aggregate lane choice data. To address this, we apply inverse optimization to estimate the preference distribution such that the resulting equilibrium flow closely matches the observed data. With the calibrated model, we compute the Wardrop equilibrium and determine the optimal toll price to achieve various policy goals. We then compare the computed toll price with the actual prices from the 2021 operations. 

\subsection{Data description} The Metropolitan Transportation Commission in California started the conversion of the existing HOV lanes to HOT lanes on the I-880 highway in 2019. The HOT lanes run from Hegenberger Road to Dixon Landing Road in the
southbound direction and from Dixon Landing Road
to Lewelling Boulevard in the northbound direction. This stretch of highway is partitioned into multiple segments. The toll price is charged for using the HOT lane on each segment from 5 am to 8 pm on each workday, and the toll is updated every 5 min. Vehicles with carpool size of $3$ can use the HOT lane for free, vehicles with carpool size of $2$ pay half of the toll price, and vehicles with a single person pay the full price.

\begin{figure}[htp]
    \centering
    \includegraphics[width=0.3\linewidth]{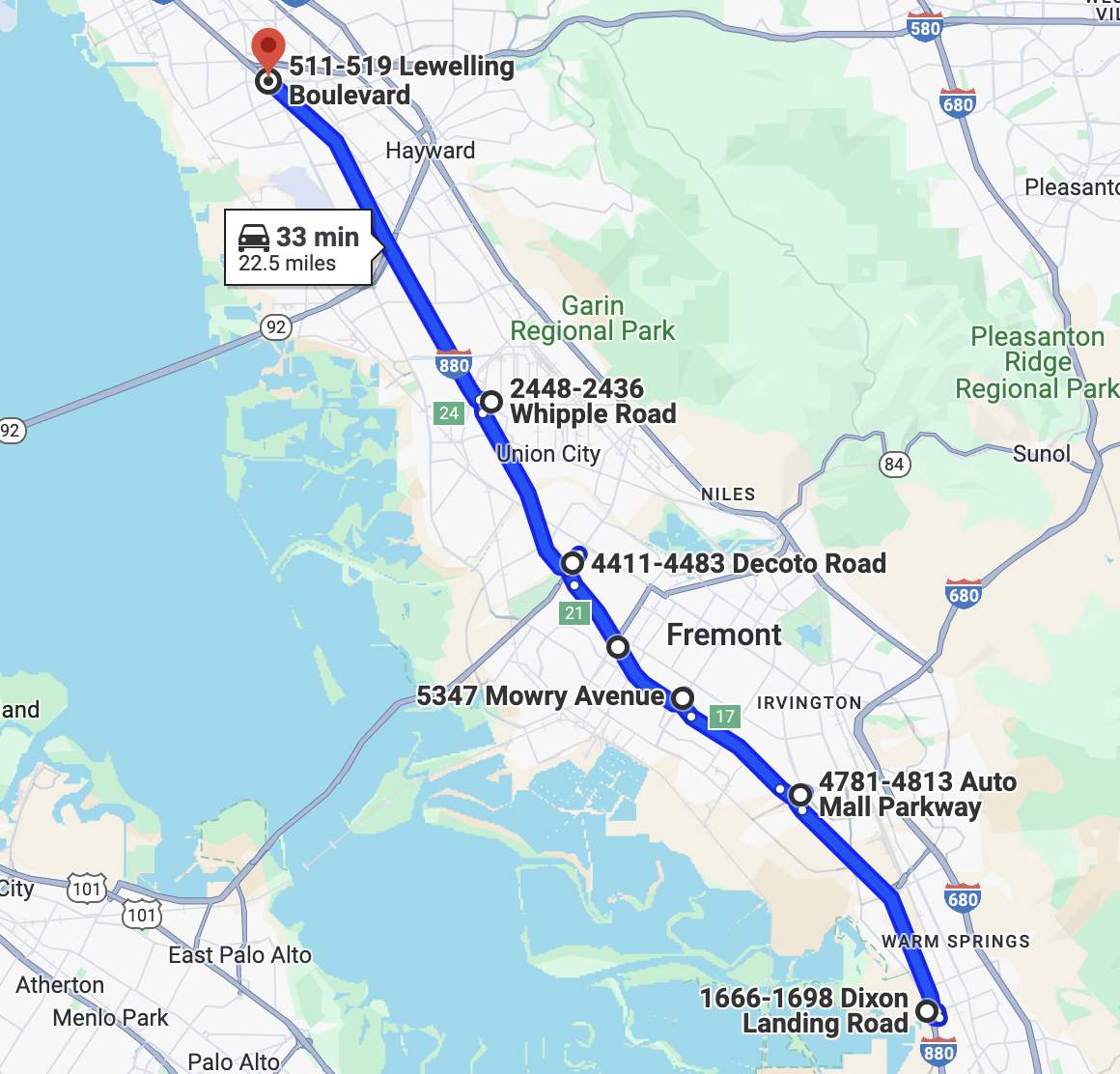}
    \caption{Interstate 880 (I-880) Highway from Dixon Landing Rd to Lewelling Blvd (Highlighted in Blue). The nodes separating the segments are marked in bold text.}
    \label{fig:map}
\end{figure}

We calibrate our multi-segment model using data collected from the Northbound of I-880 between the Dixon Landing Rd and Lewelling Blvd. The total distance is 22 miles and the highway has three ordinary lanes and one HOT lane. The highway is partitioned into five segments with separation nodes named as the Auto Mall Pkwy, Mowry Ave, Decoto Rd, Whipple Rd, and Hesperian Blvd, see Fig. \ref{fig:map}.  The distance of these segments are 5.75 miles, 3.17 miles, 3.46 miles, 2.11 miles, and 7.16 miles, respectively. 

The California Department of Transportation have installed hundreds of sensors along the I-880 highway. Each sensor measures the vehicle flow data and average speed data at the 5-minute level. We obtain the data from \cite{pems} for all workdays between March 1st 2021 and August 31st 2021. We number these workdays sequentially as $\day \in [\Day] := 1, \dots, \Day$, where $\Day$ is the total number of workdays. For each workday $\day$, we aggregate those 5-minute vehicle flows into per-hour vehicle flows of each sensor. We take the average of vehicle speed across all 5-minute intervals of each hour to infer the per-hour average speed. We divide the distance between two adjacent sensors with the average speed to compute the average travel time between each pair of adjacent sensors during each hour of the day $\day$. We number the hours from 5 am to 8 pm as $\time \in [\Horizon] := 1, \dots, \Horizon$, where $\Horizon = 15$ is the total number of HOT operation hours of each workday.

For each road segment $\segnum$ between Auto Mall Pkwy and Hesperian Blvd, we identify the list of all sensors covering this road segment. We sum up the average travel time across all these sensors to obtain the latency of ordinary lanes $\costhato{,\segnum}^{\time, \day}$ and HOT lanes $\costhath{,\segnum}^{\time, \day}$ of this road segment for each hour $\time$ of each day $\day$. Additionally, we take an average of vehicle flows across all these sensors to obtain the observed vehicle flows for ordinary lanes $\multisegflowohat{\segnum}^{\time, \day}$ and HOT lanes $\multisegflowhothat{\segnum}^{\time, \day}$ of this road segment for each hour $\time$ of each day $\day$.

\begin{figure}[htb]
    \centering
    \includegraphics[width=0.5\linewidth]{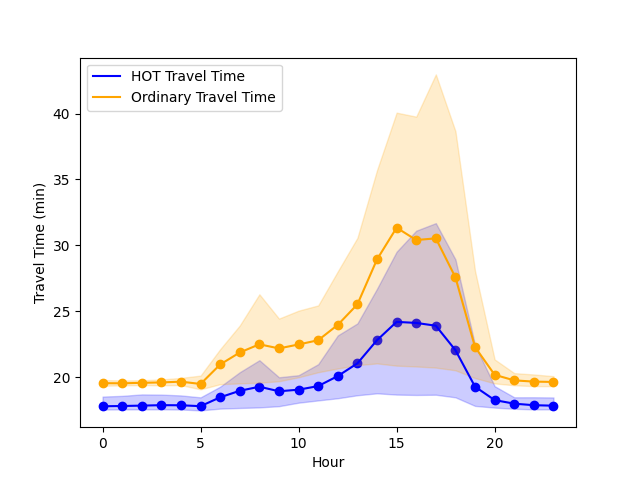}
    \caption{Average travel time (min) and 95\% confidence interval of the HOT and ordinary lanes from Auto Mall Pkwy to Hesperian Blvd in each hour of a workday.}
    \label{fig:travel-time}
\end{figure}

Figure \ref{fig:travel-time} illustrates the travel time from Auto Mall Pkwy to Hesperian Blvd in different hours of a day. The dotted lines are the mean travel time on ordinary lanes and HOT lanes in each hour of the day averaged across all days, while the shaded regions are the corresponding 95\% confidence intervals for each hour. Ordinary lanes have a uniformly higher travel time than HOT lanes in all hours. Particularly, in the afternoon hours (i.e. 2-6 pm), ordinary lanes can reach 33\% higher travel time compared to the ordinary lanes on average.

We requested from Caltrans the toll price data for each road segment for each 5 min time interval, and the daily number of vehicles at each occupancy level using the HOT lanes for the entire I-880 highway.\footnote{Due to privacy concern, the occupancy level data shared with us is aggregated across all segments and all HOT operation hours for each day.} Using this data, we compute the hourly averaged toll price of each segment and the fraction of vehicles taking the HOT lanes with occupancy level $\occu$ on each day $\day$, denoted as $\occuratio{\occu}{\day}$. 

\subsection{Model calibration}
\paragraph{Latency functions.}
We estimate the latency function of the ordinary and HOT lanes of each road segment based on the Bureau of Public Roads (BPR) function (\citet{bpr1964traffic}). Given the vehicle flow $\multisegflow$ and the HOT capacity $\hotcap=0.25$ (one lane out of 4 is HOT in the current design), the BPR function can be written as follows:
{\small
\begin{align*}
    \costo{}(\multisegflow^{}) =& T_f^{} \cdot \left[ 1 + \left( \eta \cdot \frac{\multisegflowo{}}{(1 - \hotcap) V^{}}\right)^b\right],\\
    \costh{}(\multisegflow^{}) =& T_f^{} \cdot \left[ 1 + \left( \eta \cdot \frac{\multisegflowhot{}}{\hotcap V^{}}\right)^b\right],
\end{align*}
}
where $\eta$ and $b$ are the BPR coefficients, $T_f^{}$ is the free flow travel time, and $V^{}$ is the road capacity. We set $b = 4.0$ following \cite{branston1976link}, and estimate parameters $T_f$ and $\frac{\eta}{V}$ of each segment using the flow data $\left\{\multisegflowohat{\segnum}^{\time, \day}, \multisegflowhothat{\segnum}^{\time, \day} \right\}_{\segnum \in [\segtotal], \time \in [\Horizon], \day \in [\Day]}$ and driving time$\left\{\costhato{,\segnum}^{\time, \day}, \costhath{,\segnum}^{\time, \day} \right\}_{\segnum \in [\segtotal], \time \in [\Horizon], \day \in [\Day]}$ via linear regression. 

\paragraph{Calibration of preference distribution.}
To compute the equilibrium strategy distribution, we need to estimate the population demand for each entrance and exit pair, and estimate the agents' preference distribution. To ensure the tractability of the demand and preference distribution estimates, we create a grid for each preference parameter $\vot$ and $\{\disu\}_{m \in [M]}$ with evenly spaced intervals. Then, the entire preference parameter vector space is partitioned into equally sized subsets. We denote the set of all partitioned preference sets as $K$ with generic member $k$, and assume that the preference distribution in each subset $k$ is uniform. This can be viewed as an approximation of the original probability density function of the preference parameters: As the partition of preference parameter space becomes finer, the approximation becomes closer to the original density function. 

For each pair of entrance and exit nodes $(i, j)$ and each hour \(\time\), we estimate $\demandgrid{ij, \time}{k}$ as the mass of agents with preference parameters in each subset \(k\). This estimate corresponds to the multiplication of the total agent demand for the population \((i, j)\) at time \(\time\) and the fraction of agents with preference parameters in subset \(k\). Our estimate captures the variation of agent demand and preference distribution across different entrance and exit nodes and times of the day. We assume the agents' demand for each hour is the same across all workdays. 

We estimate $\demandgrid{}{} := \left(\demandgrid{ij, \time}{k} \right)_{i \leq j \in [\segtotal], \time \in [\Horizon], k \in [K]}$ using inverse optimization. Given data on toll prices $\left\{\toll_{\segnum, \occu}^{\time, \day} \right\}_{\segnum \in [\segtotal]}$ and driving time $\left\{\costhato{,\segnum}^{\time, \day}, \costhath{,\segnum}^{\time, \day} \right\}_{\segnum \in [\segtotal]}$ for each hour $\time$ of each day $\day$, we compute the best response strategy profile $\multisegstrategyeq{\time, \day}$ of all agents for hour $\time$ and day $\day$. With the estimated demand $\left(\demandgrid{ij, \time}{k} \right)_{i \leq j \in [\segtotal], k \in [K]}$ for hour $\time$, we obtain the best response strategy distribution $\left(\multisegstrategydiseq{\action}{\time, \day}(\demandgrid{}{}) \right)_{\action \in \actionset}$ following analysis in Sec. \ref{sec:multiseg}, and the induced vehicle flow $\multisegfloweq{}{\time, \day}(\demandgrid{}{}) := \left\{\multisegflowoeq{\segnum}{\time, \day}(\demandgrid{}{}), \multisegflowhoteq{\segnum}{\time, \day}(\demandgrid{}{}) \right\}_{\segnum \in [\segtotal]}$ based on \eqref{eq:multiseg-flow}. Additionally, we compute the induced best response vehicle flow of each occupancy level $\multisegflowoccu{*\day}{}(\demandgrid{}{}) := \left(\multisegflowoccu{*\day}{\occu}(\demandgrid{}{}) \right)_{\occu \in [\occumax]}$ as follows: $\forall \occu \in [\occumax]$,
{\small
\begin{align} \label{eq:multiseg-flow-ratio}
    &\multisegflowoccu{*\day}{\occu}(\demandgrid{}{}) = \sum_{\time \in [\Horizon]} \sum_{i \leq j \in [\segtotal]} \demand{ij, \time}{} \nonumber \\
    &\cdot \left(\sum_{\action \in \actionset^{ij}} \frac{1}{\actionoccu} \multisegstrategydiseq{a}{ij, \time, \day}(\demandgrid{}{}) \mathds{1}_{\actionseg{\segnum} = \mathrm{h}, \actionoccu = \occu} \right).
\end{align}
}

We estimate $\demandgrid{}{}$ to be the demand vector such that the induced best response vehicle flow is close to the observed flows $\{\multisegflowohat{\segnum}^{\time, \day}, \multisegflowhothat{\segnum}^{\time, \day}\}_{\segnum \in [\segtotal], \time \in [\Horizon], \day \in [\Day]}$ and the induced fraction of vehicles taking each occupancy level is close to $\left(\occuratio{\occu}{\day} \right)_{\occu \in [\occumax], \day \in [\Day]}$. We formulate the estimation problem as the following convex optimization program: 
{\small
\begin{align*}
    \min_{\demandgrid{}{}}&\ \sum_{\day \in [\Day]} \sum_{\day \in [\Day]} \sum_{\occu \in [\occumax]} \left(\occuratio{\occu}{\day} \left(\sum_{\occu' \in [\occumax]} \multisegflowoccu{*\day}{\occu'}(\demandgrid{}{})\right) - \multisegflowoccu{*\day}{\occu}(\demandgrid{}{}) \right)^2 \\
    +& \sum_{\time \in [\Horizon]} \sum_{\segnum \in [\segtotal]} \left(\left(\multisegflowoeq{\segnum}{\time, \day}(\demandgrid{}{}) - \multisegflowohat{\segnum}^{\time, \day} \right)^2 + \left(\multisegflowhoteq{\segnum}{\time, \day}(\demandgrid{}{}) - \multisegflowhothat{\segnum}^{\time, \day} \right)^2\right) .
\end{align*}
}

\begin{figure}[htb]
    \centering
    \includegraphics[width=0.6\linewidth]{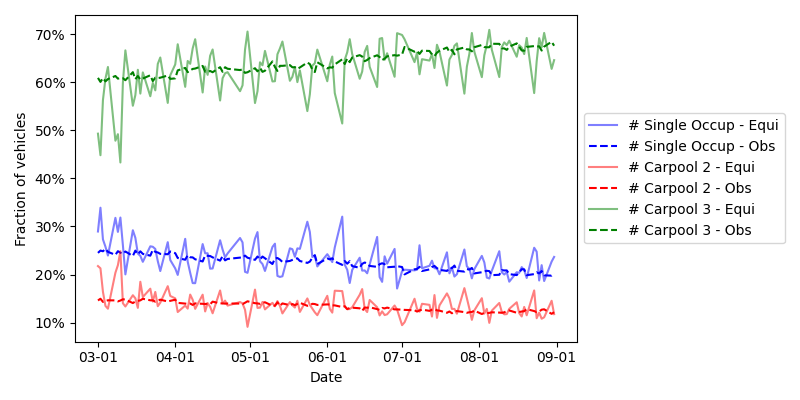}
    \caption{Fractions of agents on HOT lanes taking each occupancy level per day.}
    \label{fig:daily-ratio}
\end{figure}

We use the calibrated demand to compute the number of agents taking each occupancy level and compare it with the actual numbers. In Fig. \ref{fig:daily-ratio}, the dotted lines show the observed daily fraction of agents on HOT lanes for each occupancy level, while the curved lines show the equilibrium daily ratio of agents for each occupancy level based on calibrated demand and induced equilibrium. The close alignment of curved lines and dotted lines demonstrates that our calibrated demand closely matches the actual demand.

\subsection{Optimal design of toll price and HOT capacity at 5-6pm}
Based on the calibrated model, we compute the optimal toll price and HOT capacity allocation for 5-6 pm. We consider each of the following four objectives in equilibrium: 

\begin{enumerate}
    \item The total agent travel time.
    {\small
    \begin{align} \label{eq:opt-total-travel-time}
        &\objcongestion(\toll, \hotcap) := \sum_{i \leq j \in [\segtotal]} \demand{ij}{} \left[\sum_{\action \in \actionset^{ij}} \multisegstrategydiseq{\action}{ij} \right. \nonumber\\
        &\cdot \left. \left(\sum_{\segnum = i}^j \costo{,\segnum}(\multisegflowoeq{\segnum}{}, 1-\hotcap) \mathds{1}_{\actionseg{\segnum} = \mathrm{o}} + \costh{,\segnum}(\multisegflowhoteq{\segnum}{}, \hotcap) \mathds{1}_{\actionseg{\segnum} = \mathrm{h}} \right) \right].
    \end{align}
    }
    \item The total vehicle driving time. The total vehicle driving time differs from the total agent travel time. Specifically, for every minute a vehicle with \(\occu \in [\occumax]\) agents spends on the highway, the total vehicle driving time counts this as 1 minute, whereas the agent travel time counts it as \(\occu\) minutes.
    {\small
    \begin{align} \label{eq:opt-total-emission}
        &\objemission(\tau, \rho) := \sum_{i \leq j \in [\segtotal]} \demand{ij}{} \left[\sum_{\action \in \actionset^{ij}} \multisegstrategydiseq{\action}{ij} \frac{1}{\actionoccu} \right. \nonumber \\
        &\cdot \left. \left(\sum_{\segnum = i}^j \costo{,\segnum}(\multisegflowoeq{\segnum}{}, 1-\hotcap) \mathds{1}_{\actionseg{\segnum} = \mathrm{o}} + \costh{,\segnum}(\multisegflowhoteq{\segnum}{}, \hotcap) \mathds{1}_{\actionseg{\segnum} = \mathrm{h}} \right) \right].
    \end{align}
    }
    \item Total revenue: The total toll prices paid by all agents.
    {\small
    \begin{equation} \label{eq:opt-total-revenue}
        \objrevenue(\tau, \rho) := \sum_{i \leq j \in [\segtotal]} \demand{ij}{} \left[\sum_{\action \in \actionset^{ij}} \multisegstrategydiseq{\action}{ij} \frac{1}{\actionoccu} \left(\sum_{\segnum = i}^j \toll_{\segnum, \actionoccu} \mathds{1}_{\actionseg{\segnum} = \mathrm{h}} \right) \right].
    \end{equation}
    }
    \item Total cost: The total cost (driving time, toll price and carpool disutility) incurred by all agents incorporating their heterogeneous preferences.
    {\small
    \begin{align} \label{eq:opt-total-cost}
        &\objutility(\tau, \rho) := \sum_{i \leq j \in [\segtotal]} \demand{ij}{} \left[\sum_{\action \in \actionset^{ij}} \int_{\disu} \int_{\vot} \costa(\multisegstrategydiseq{}{}, \vot, \disu) \right. \nonumber\\
        &\cdot \left. \mathds{1}_{ \multisegstrategyeq{ij}(\vot, \disu)=a} \pdf^{ij}(\vot, \disu) d\vot d\disu \right].
    \end{align}
    }
\end{enumerate}

The problems \eqref{eq:opt-total-travel-time} -- \eqref{eq:opt-total-cost} compute the HOT toll price for each segment that optimizes the objective function based on the induced equilibrium strategy. As such, these are mathematical programs with equilibrium constraints (MPEC). Furthermore, since the toll is only applied to the HOT lane, these problems can be seen as a generalization of optimal tolling with support constraints, which has been proven to be NP-hard (\cite{hoefer2008taxing, harks2015computing}). We adopt the enumeration algorithm for the optimal design of toll price and HOT capacity. In particular, we set the toll price on each road segment within the range of $0$ and $7$ dollars discretized by $\$0.5$. We also choose the set $\rho \in\{ \frac{1}{4}, \frac{2}{4}, \frac{3}{4}\}$ since the highway has four lanes. For each pair of $(\tau, \rho)$, we compute the equilibrium strategies, and the corresponding objective function value. We choose the toll price and capacity fraction with the optimal value. 

Table \ref{tab:toll-capacity-design} summarizes our result. The first row shows the current HOT capacity and the average toll price. The remaining rows in Table \ref{tab:toll-capacity-design} shows the optimal toll prices on each road segment for agent travel time minimization, vehicle driving time minimization, revenue maximization, and cost minimization, under different HOT capacities.

When the HOT capacity is 0.25 (the setting in practice), the optimal toll prices for agent travel time, vehicle driving time, and cost minimization are lower than the current prices on Auto Mall Pkwy, Mowry Ave, Decoto Rd, and Whipple Rd. On Hesperian Blvd, they match the current average toll price. For revenue maximization, the optimal toll prices are lower than the current average prices on all five road segments. As HOT capacity increases, the optimal toll prices for these four objectives may vary, either increasing or decreasing, depending on each road segment.

Note that the optimal toll prices for agent travel time, vehicle driving time, and cost minimization are higher on Hesperian Blvd, and lower on other segments. We remark that it aligns with the fact that the demand volume of agents is also higher on Hesperian Blvd and lower on other segments. 
Given the convex nature of the latency function, a large demand volume on a road segment means that the reduction in travel time achieved by incentivizing agents to carpool becomes more significant compared to segments with smaller demand volumes. Therefore, setting a high toll price on road segments with large demand is more effective in incentivizing carpool, which in turn reduces the agent travel time, vehicle driving time, and cost of agents. 

\begin{figure}[htb]
    \centering
    \begin{subfigure}{0.33\linewidth}
        \centering
        \includegraphics[width=\linewidth]{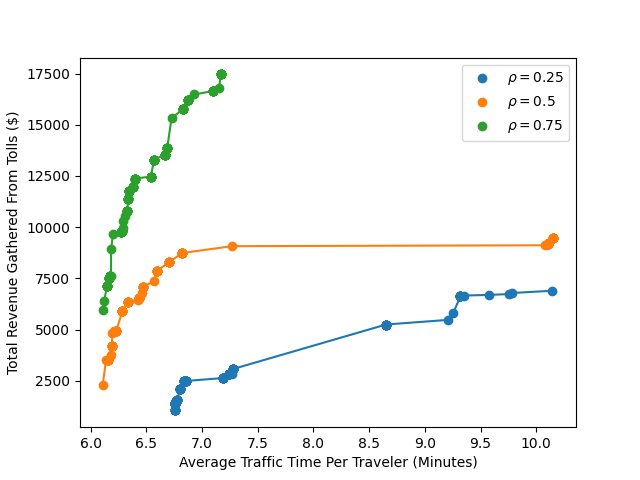}
        \caption{Agent Time}
        \label{subfig:pareto-congestion}
    \end{subfigure}%
    \begin{subfigure}{0.33\linewidth}
        \centering
        \includegraphics[width=\linewidth]{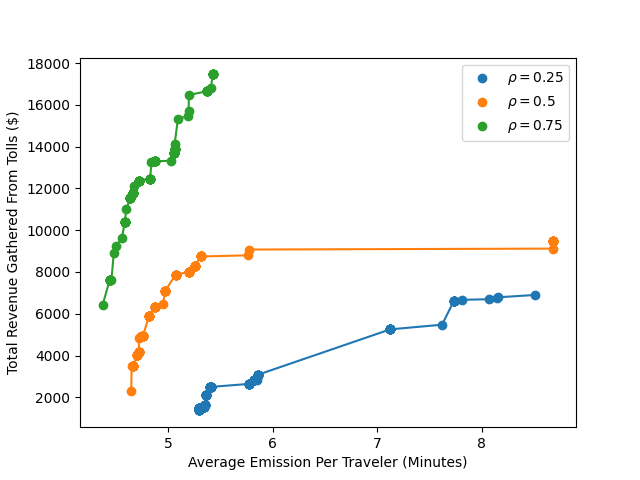}
        \caption{Vehicle Time}
        \label{subfig:pareto-emission}
    \end{subfigure}%
    \begin{subfigure}{0.33\linewidth}
        \centering
        \includegraphics[width=\linewidth]{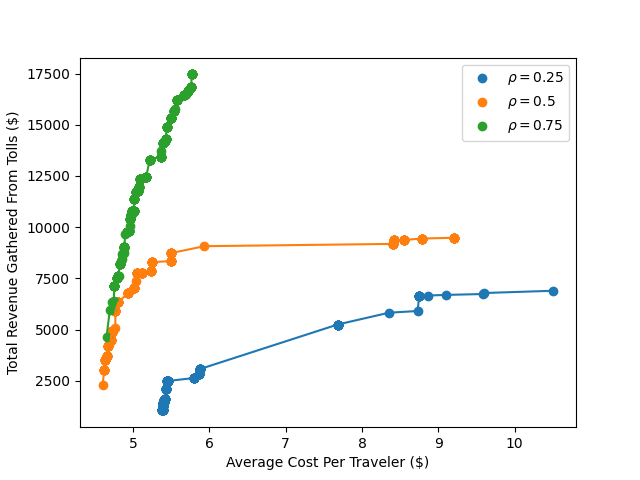}
        \caption{Cost}
        \label{subfig:pareto-utility-cost}
    \end{subfigure}
    \caption{Pareto fronts of agent travel time, vehicle driving time, and cost w.r.t revenue for 5-6 pm on all road segments. The blue, orange, and green curves show the maximum attainable revenue at each value of agent travel time, vehicle driving time, and cost for HOT capacity of 0.25, 0.5, and 0.75 respectively.}
    \label{fig:pareto}
\end{figure}

However, charging a high toll price can lead agents to either carpool or take the ordinary lane, leaving fewer agents willing to pay the toll. Consequently, the optimal toll price that maximizes the revenue is often lower than the ones associated with the other three objectives. This creates a trade-off between revenue maximization and the minimization of agent travel time, vehicle driving time, and costs. This tradeoff is illustrated in the Pareto front in Fig. \ref{fig:pareto}. The blue, orange, and green curves in Fig. \ref{fig:pareto} show the maximum attainable revenue at each value of agent travel time, vehicle driving time, and cost for HOT capacity of 0.25, 0.5, and 0.75, respectively.\footnote{While the optimal toll design of a high HOT capacity dominates the optimal toll design of lower HOT capacities, it does not mean that setting more lanes as the HOT lane is better in reality. We need to take other aspects into considerations, for example the accessibility and equity of agents with different incomes.}


\begin{table}[htb]
    \centering
    \begin{tabular}{|c|c|c|c|c|c|c|}
        \hline
        \multirow{2}{*}{HOT Capacity} & \multirow{2}{*}{Objective} & \multicolumn{5}{c|}{Toll Prices} \\
         & & Auto Mall & Mowry & Decoto & Whipple & Hesperian\\
        \hline
        0.25 & Current Prices & $\$1.1$ & $\$2.2$ & $\$2.5$ & $\$4.0$ & $\$5.0$\\
        \hline
        \multirow{4}{*}{0.25} & Agent Time Minimization & $\$0.5$ & $\$1.5$ & $\$0.5$ & $\$0.5$ & $\$5.0$\\
         & Vehicle Time Minimization & $\$0.5$ & $\$1.5$ & $\$0.5$ & $\$0.5$ & $\$4.0$\\
         & Revenue Minimization & $\$0.0$ & $\$1.0$ & $\$1.0$ & $\$0.5$ & $\$1.5$\\
         & Cost Minimization & $\$0.5$ & $\$1.5$ & $\$0.5$ & $\$0.5$ & $\$5.0$\\
        \hline
        \multirow{4}{*}{0.50} & Agent Time Minimization & $\$0.0$ & $\$0.5$ & $\$0.0$ & $\$0.0$ & $\$5.0$\\
         & Vehicle Time Minimization & $\$0.0$ & $\$0.5$ & $\$0.0$ & $\$0.0$ & $\$5.0$\\
         & Revenue Minimization & $\$0.5$ & $\$1.5$ & $\$1.5$ & $\$1.5$ & $\$1.0$\\
         & Cost Minimization & $\$0.0$ & $\$0.5$ & $\$0.0$ & $\$0.0$ & $\$5.0$\\
        \hline
        \multirow{4}{*}{0.75} & Agent Time Minimization & $\$0.0$ & $\$0.5$ & $\$0.0$ & $\$0.0$ & $\$5.0$\\
         & Vehicle Time Minimization & $\$0.0$ & $\$1.5$ & $\$0.0$ & $\$0.0$ & $\$5.0$\\
         & Revenue Minimization & $\$0.5$ & $\$2.5$ & $\$4.5$ & $\$1.5$ & $\$4.5$\\
         & Cost Minimization & $\$0.0$ & $\$0.0$ & $\$0.0$ & $\$0.0$ & $\$5.0$\\
        \hline
    \end{tabular}
    \caption{Optimal toll prices and HOT capacity design for 5-6 pm on all five road segments}
    \label{tab:toll-capacity-design}
\end{table}

\subsection{Hourly optimal toll pricing}
\begin{figure}[htb]
    \centering
    \begin{tabular}{|c|cccc|}
        \hline
        Auto Mall & \begin{subfigure}{0.2\linewidth}
            \centering
            \includegraphics[width=\linewidth]{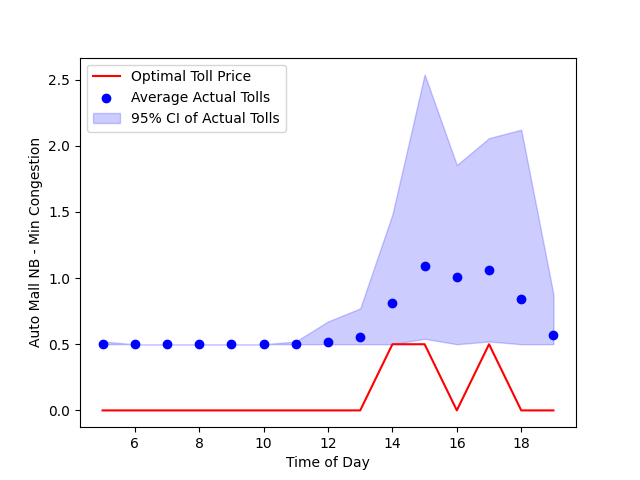}
            \caption{Agent Time}
            \label{subfig:dynamic-toll-auto-mall-congestion}
        \end{subfigure} & \begin{subfigure}{0.2\linewidth}
            \centering
            \includegraphics[width=\linewidth]{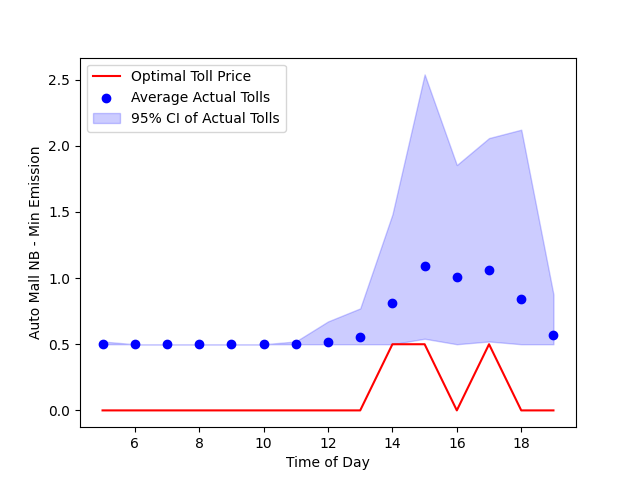}
            \caption{Vehicle Time}
            \label{subfig:dynamic-toll-auto-mall-emission}
        \end{subfigure} & \begin{subfigure}{0.2\linewidth}
            \centering
            \includegraphics[width=\linewidth]{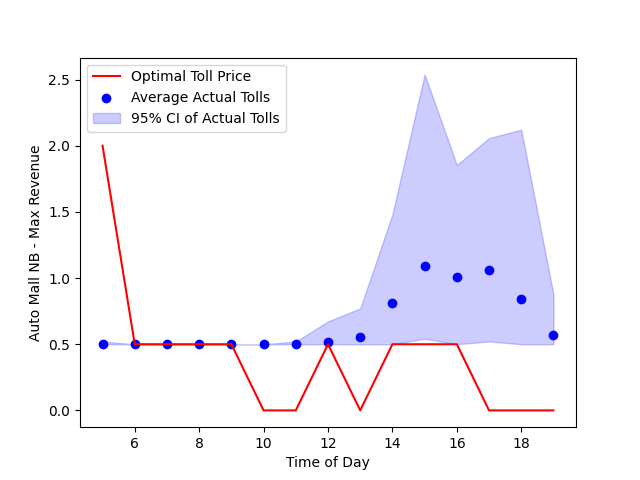}
            \caption{Revenue}
            \label{subfig:dynamic-toll-auto-mall-revenue}
        \end{subfigure} & \begin{subfigure}{0.2\linewidth}
            \centering
            \includegraphics[width=\linewidth]{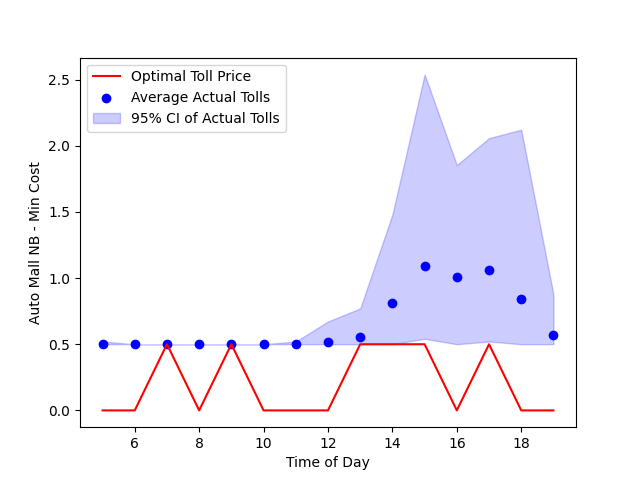}
            \caption{Cost}
            \label{subfig:dynamic-toll-auto-mall-utility-cost}
        \end{subfigure}\\
        \hline
        Mowry & \begin{subfigure}{0.2\linewidth}
            \centering
            \includegraphics[width=\linewidth]{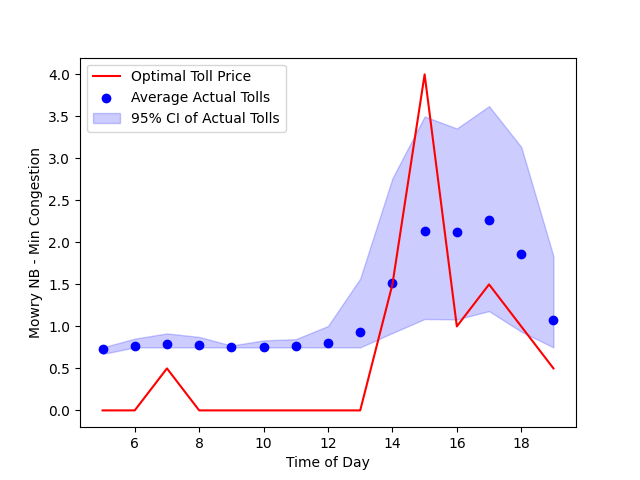}
            \caption{Agent Time}
            \label{subfig:dynamic-toll-mowry-congestion}
        \end{subfigure} & \begin{subfigure}{0.2\linewidth}
            \centering
            \includegraphics[width=\linewidth]{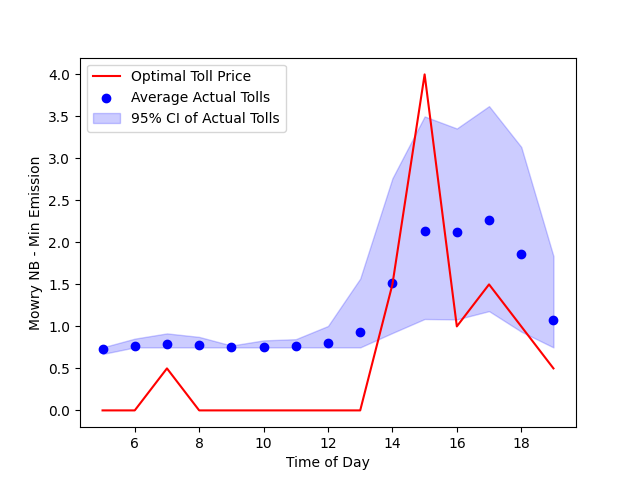}
            \caption{Vehicle Time}
            \label{subfig:dynamic-toll-mowry-emission}
        \end{subfigure} & \begin{subfigure}{0.2\linewidth}
            \centering
            \includegraphics[width=\linewidth]{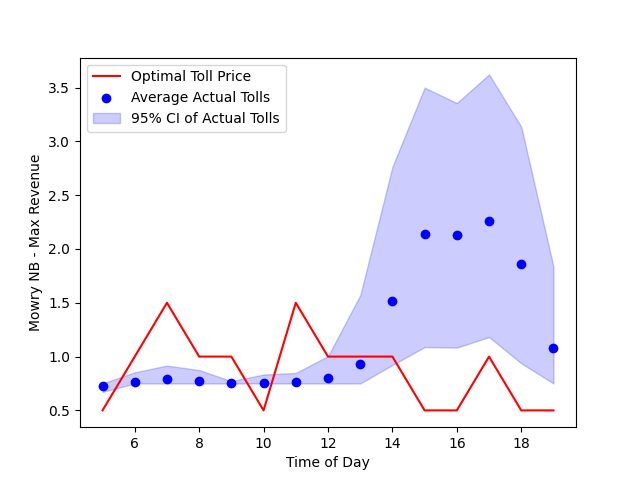}
            \caption{Revenue}
            \label{subfig:dynamic-toll-mowry-revenue}
        \end{subfigure} & \begin{subfigure}{0.2\linewidth}
            \centering
            \includegraphics[width=\linewidth]{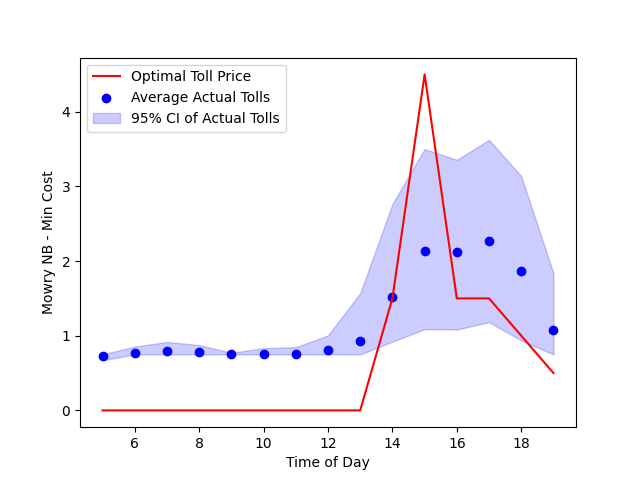}
            \caption{Cost}
            \label{subfig:dynamic-toll-mowry-utility-cost}
        \end{subfigure}\\
        \hline
        Decoto & \begin{subfigure}{0.2\linewidth}
            \centering
            \includegraphics[width=\linewidth]{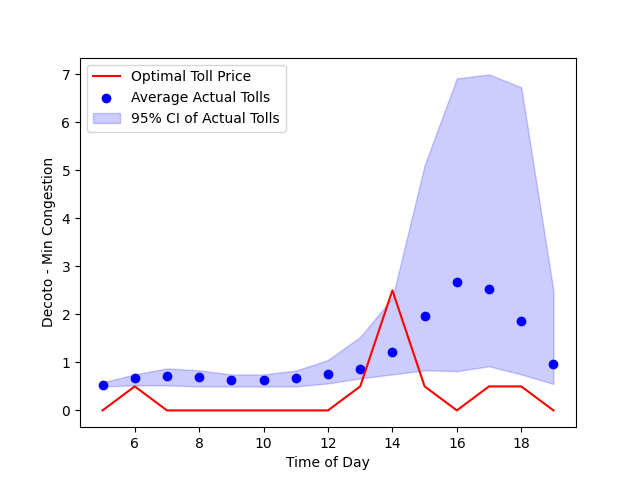}
            \caption{Agent Time}
            \label{subfig:dynamic-toll-decoto-congestion}
        \end{subfigure} & \begin{subfigure}{0.2\linewidth}
            \centering
            \includegraphics[width=\linewidth]{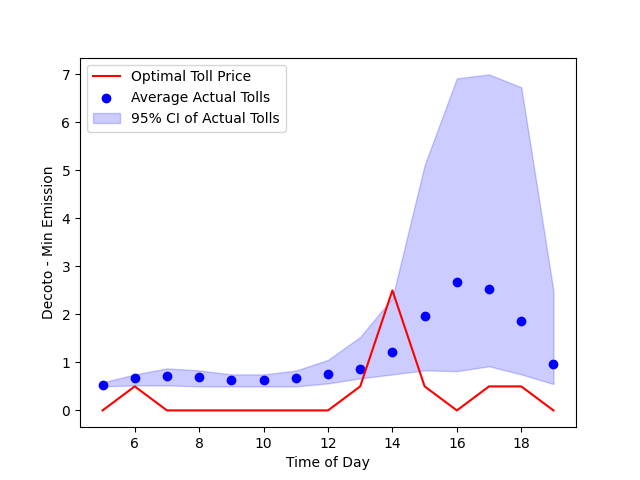}
            \caption{Vehicle Time}
            \label{subfig:dynamic-toll-decoto-emission}
        \end{subfigure} & \begin{subfigure}{0.2\linewidth}
            \centering
            \includegraphics[width=\linewidth]{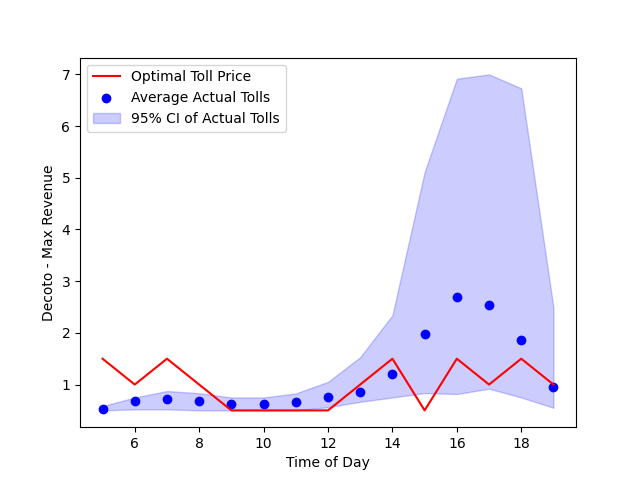}
            \caption{Revenue}
            \label{subfig:dynamic-toll-decoto-revenue}
        \end{subfigure} & \begin{subfigure}{0.2\linewidth}
            \centering
            \includegraphics[width=\linewidth]{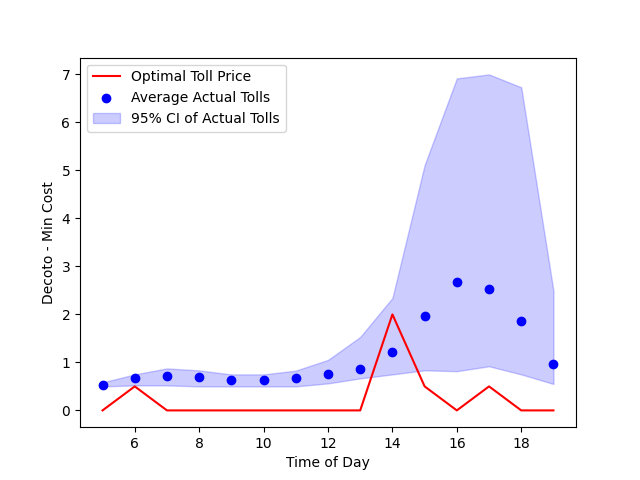}
            \caption{Cost}
            \label{subfig:dynamic-toll-decoto-utility-cost}
        \end{subfigure}\\
        \hline
        Whipple & \begin{subfigure}{0.2\linewidth}
            \centering
            \includegraphics[width=\linewidth]{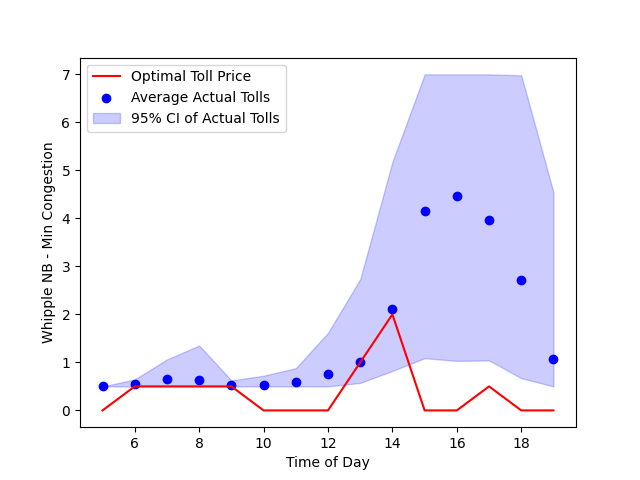}
            \caption{Agent Time}
            \label{subfig:dynamic-toll-whipple-congestion}
        \end{subfigure} & \begin{subfigure}{0.2\linewidth}
            \centering
            \includegraphics[width=\linewidth]{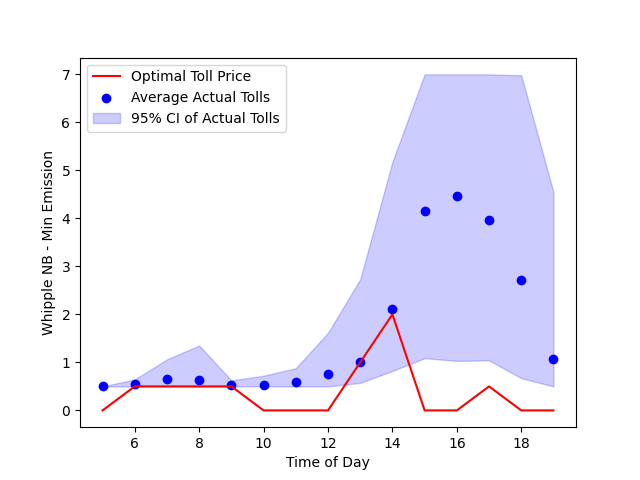}
            \caption{Vehicle Time}
            \label{subfig:dynamic-toll-whipple-emission}
        \end{subfigure} & \begin{subfigure}{0.2\linewidth}
            \centering
            \includegraphics[width=\linewidth]{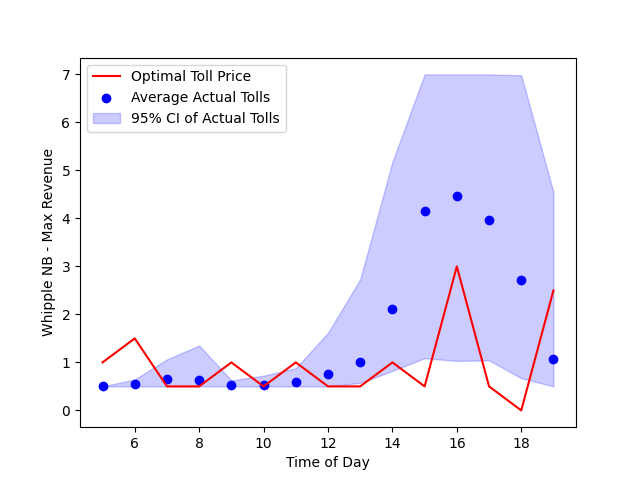}
            \caption{Revenue}
            \label{subfig:dynamic-toll-whipple-revenue}
        \end{subfigure} & \begin{subfigure}{0.2\linewidth}
            \centering
            \includegraphics[width=\linewidth]{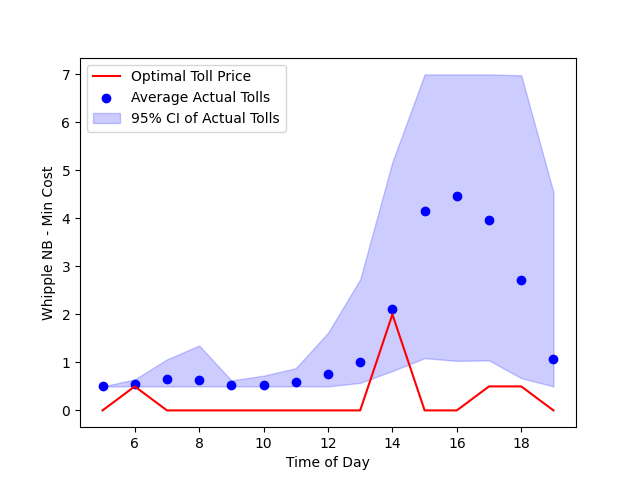}
            \caption{Cost}
            \label{subfig:dynamic-toll-whipple-utility-cost}
        \end{subfigure}\\
        \hline
        Hesperian & \begin{subfigure}{0.2\linewidth}
            \centering
            \includegraphics[width=\linewidth]{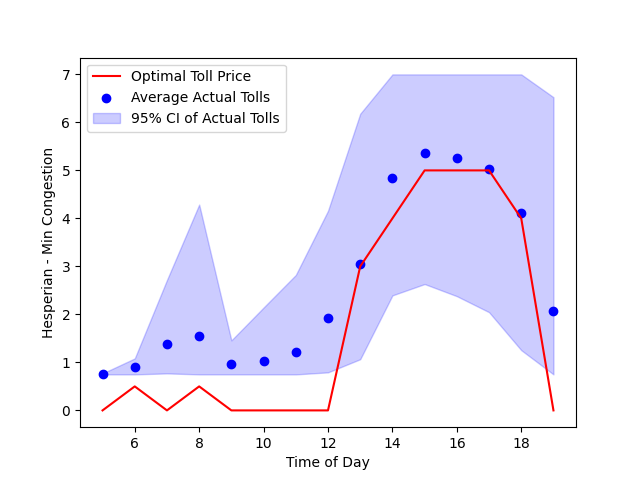}
            \caption{Agent Time}
            \label{subfig:dynamic-toll-hesperian-congestion}
        \end{subfigure} & \begin{subfigure}{0.2\linewidth}
            \centering
            \includegraphics[width=\linewidth]{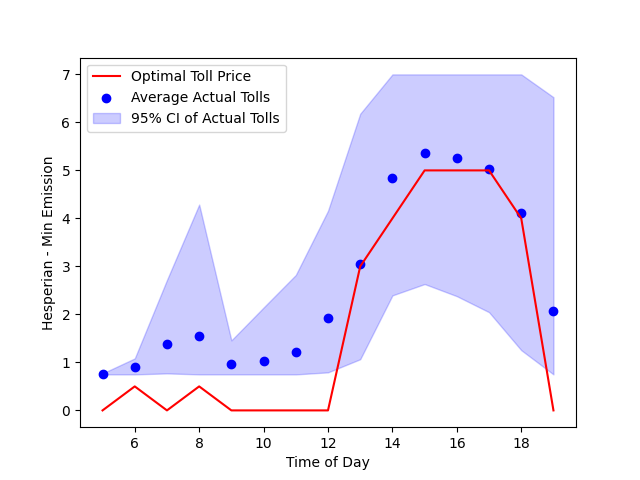}
            \caption{Vehicle Time}
            \label{subfig:dynamic-toll-hesperian-emission}
        \end{subfigure} & \begin{subfigure}{0.2\linewidth}
            \centering
            \includegraphics[width=\linewidth]{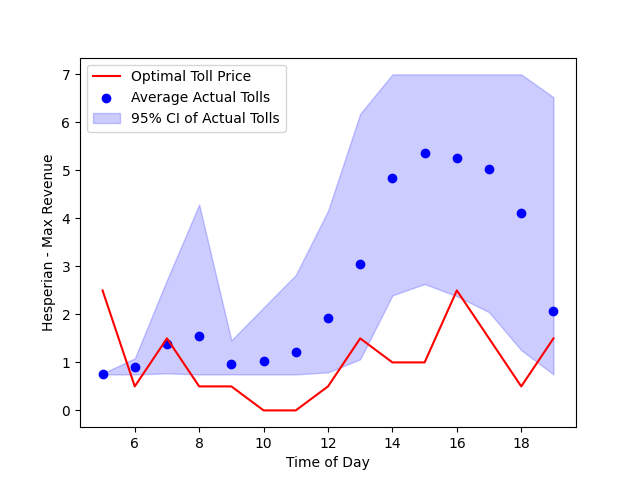}
            \caption{Revenue}
            \label{subfig:dynamic-toll-hesperian-revenue}
        \end{subfigure} & \begin{subfigure}{0.2\linewidth}
            \centering
            \includegraphics[width=\linewidth]{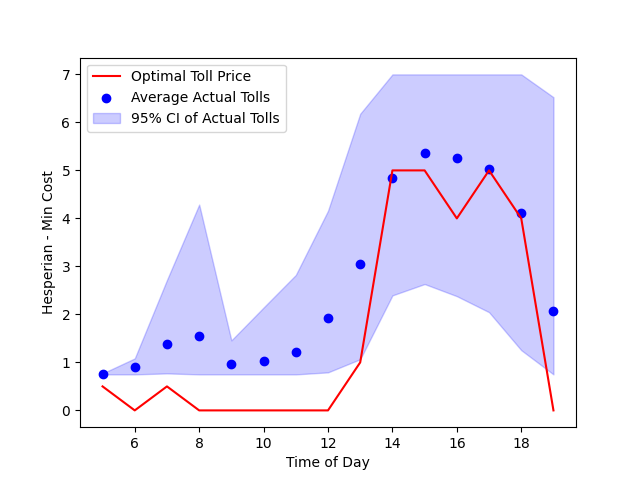}
            \caption{Cost}
            \label{subfig:dynamic-toll-hesperian-utility-cost}
        \end{subfigure}\\
        \hline
    \end{tabular}
    \caption{Optimal toll prices for agent travel time minimization, vehicle driving time minimization, revenue maximization, and cost minimization in each hour from 5 am to 8 pm of a workday. Red curves represent the optimal toll prices, blue dots represent the average current toll prices at each hour, while the blue shaded regions represent the 95\% confidence regions of the current toll prices, across all workdays from March 1st 2021 to August 31st 2021.}
    \label{fig:dynamic-toll}
\end{figure}

In this section, we compute the optimal hourly toll price from 5am to 8pm for each one of the four objectives. 
We set the HOT capacity as $0.25$ to match the current HOT capacity on I-880. In Figure \ref{fig:dynamic-toll}, red curves represent the optimal toll prices, blue dots represent the average current toll price recorded in the data, and the blue shaded regions represent the 95\% confidence interval of the current toll price.

For Auto Mall Parkway, the optimal toll prices for minimizing agent travel time, vehicle driving time, and costs are lower than the current prices for most hours of the day. The optimal toll prices for revenue maximization are higher in the early morning but lower in the afternoon (Figures \ref{subfig:dynamic-toll-auto-mall-congestion}-\ref{subfig:dynamic-toll-auto-mall-utility-cost}). On Mowry Ave (Figures \ref{subfig:dynamic-toll-mowry-congestion}-\ref{subfig:dynamic-toll-mowry-utility-cost}), the optimal toll prices for agent travel time minimization, vehicle driving time minimization, and cost minimization are lower than the current prices in the morning (before 12 pm) but higher during evening rush hours (around 5 pm). The optimal toll prices for revenue maximization are lower than the current prices during evening rush hours but similar in other hours. For Decoto Rd (Figures \ref{subfig:dynamic-toll-decoto-congestion}-\ref{subfig:dynamic-toll-decoto-utility-cost}) and Whipple Rd (Figures \ref{subfig:dynamic-toll-whipple-congestion}-\ref{subfig:dynamic-toll-whipple-utility-cost}), the optimal toll prices for all four objectives are lower than the current prices during evening rush hours and similar to the current prices at other times. Lastly, for Hesperian Blvd (Figures \ref{subfig:dynamic-toll-hesperian-congestion}-\ref{subfig:dynamic-toll-hesperian-utility-cost}), the optimal toll prices for agent travel time and vehicle driving time minimization are lower than the current prices during morning rush hours (6-8 am), while the optimal toll prices for revenue maximization are lower during evening rush hours.

On each road segment, the optimal toll prices for agent travel time minimization, vehicle driving time minimization, and cost minimization are almost identical at each hour. However, the optimal toll prices for revenue maximization are similar during the morning hours (before 12 pm) but significantly lower than the optimal toll prices for the other three objectives during the evening rush hours (around 5 pm), especially on Hesperian Blvd.

In the morning, when the agent demand volume is low, setting a high toll price to incentivize carpooling does not lead to a significant reduction in HOT latency. Therefore, the optimal toll prices for both the minimization of agent travel time, vehicle driving time, and costs, as well as for revenue maximization, are low during the morning hours.

On the other hand, in the evening hours, when agent demand is higher, setting a high toll price can lead to a substantial reduction in HOT latency, particularly on road segments with high demand. However, this also discourages most people from paying the toll. As a result, the optimal toll prices for revenue maximization are lower than those for minimizing agent travel time, vehicle driving time, and costs. This difference is more pronounced on road segments with high agent demand, such as Hesperian Blvd.

\begin{figure}[htb]
    \centering
    \begin{subfigure}{0.4\linewidth}
        \centering
        \includegraphics[width=\linewidth]{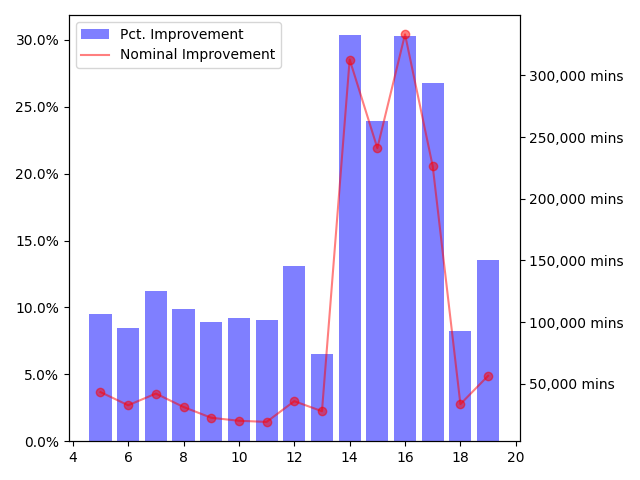}
        \caption{Agent Time}
        \label{subfig:improvement-congestion}
    \end{subfigure}%
    \begin{subfigure}{0.4\linewidth}
        \centering
        \includegraphics[width=\linewidth]{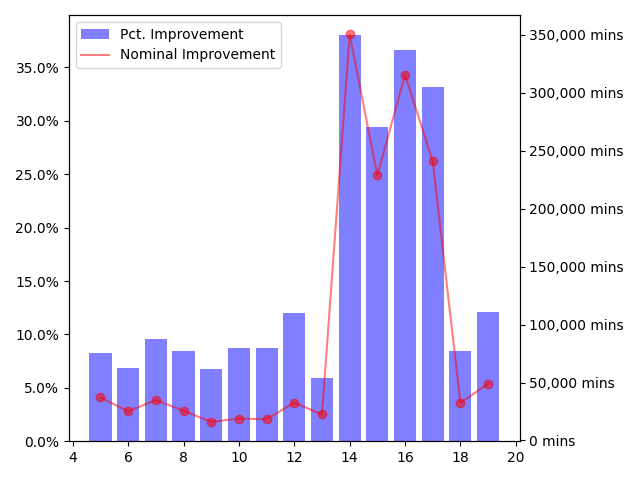}
        \caption{Vehicle Time}
        \label{subfig:improvement-emission}
    \end{subfigure}\\
    \begin{subfigure}{0.4\linewidth}
        \centering
        \includegraphics[width=\linewidth]{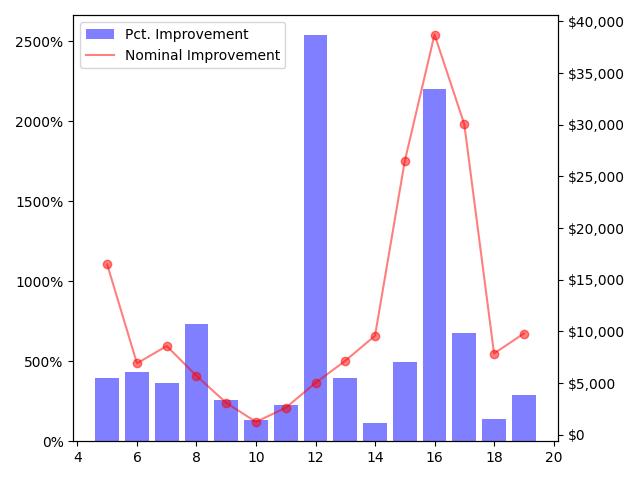}
        \caption{Revenue}
        \label{subfig:improvement-revenue}
    \end{subfigure}%
    \begin{subfigure}{0.4\linewidth}
        \centering
        \includegraphics[width=\linewidth]{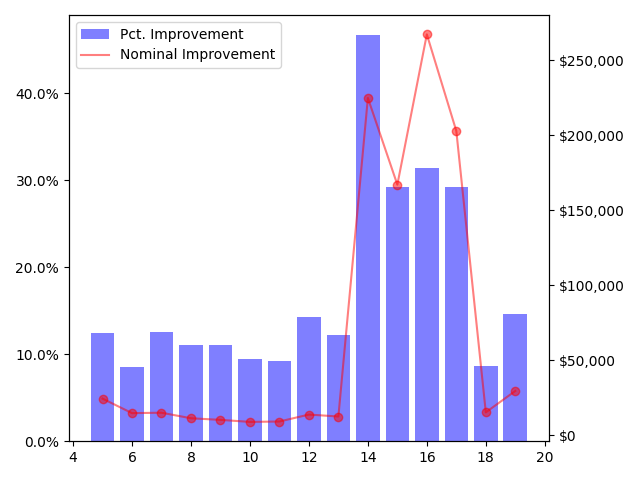}
        \caption{Cost}
        \label{subfig:improvement-utility-cost}
    \end{subfigure}
    \caption{Improvement of objectives by optimal toll price design in each hour from 5 am to 8 pm of a workday}
    \label{fig:improvement}
\end{figure}

Figure \ref{fig:improvement} illustrates the improvements achieved by implementing optimal toll prices in terms of agent travel time, vehicle driving time, revenue, and cost, compared to the current toll prices. Blue bars represent the percentage improvements, while red curves depict the numerical improvements. By using the optimal toll prices, we can achieve reductions of up to 30-40\% in agent travel time, vehicle driving time, and costs, and an increase in revenue of up to 2500\%. Specifically, optimal toll prices can lead to reductions of up to 300,000 minutes in total agent travel time, 350,000 minutes in total vehicle driving time, and \$250,000 in total costs, while increasing total revenue by up to \$40,000. The largest numerical improvements for all four objectives occur during the afternoon hours, when travel demand is high.

%% file: conclusion.tex
In this article, we examine a game-theoretic model that analyzes the lane choice of travelers with heterogeneous values of time and carpool disutilites on highways equipped with HOT lanes. For highways with a single road segment, we characterize the equilibrium strategies, and identify two qualitatively distinct equilibrium regimes that depends on the HOT lane capacity and toll price. We discuss how equilibrium strategies and latency difference of ordinary lanes and HOT lanes will change by increasing the HOT capacity or toll price. Additionally, we extend our model to highways with multiple entrance and exit nodes. We calibrate our model using the data of California Interstate highway 880 and determine the optimal capacity allocation and toll design. As a future direction of research, we will investigate the equilibrium property in a fully generalized network, and the design of HOT systems with multiple combined objectives. 

%% file: appendix.tex
\section{Proof of Results}\label{apx:proof}
\noindent\textit{Proof of Lemma \ref{lem:possibleOutcomes}.} We first show that $\costdiff(\strategydiseq{}, \hotcap) > 0$. Assume that $\costdiff(\strategydiseq{}, \hotcap) \leq 0$ for the sake of contradiction. Then, for any $(\vot, \disu)$ with $\disu > 0$, we have 
\begin{align*}
    C_{\actiono}(\strategydiseq{}, \vot, \disu) - C_{\actiont}(\strategydiseq{}, \vot, \disu) = \vot\costdiff(\strategydiseq{}, \hotcap) - \toll < 0,\\
    C_{\actiono}(\strategydiseq{}, \vot, \disu) - C_{\actionp}(\strategydiseq{}, \vot, \disu) = \vot\costdiff(\strategydiseq{}, \hotcap) - \gamma < 0.
\end{align*}
That is, all agent will choose to take the ordinary lane, and thus
\begin{align*}
    &\costdiff(\strategydiseq{}, \hotcap) = \costo{}(D, 1 - \hotcap) - \costh{}(0, \hotcap)\\ 
    =& \costo{}(D, 1 - \hotcap) - \costo{}(0, 1-\hotcap) > 0,
\end{align*}
where the second equality is due to Assumption \ref{as:cost} (b) and the last inequality is due to Assumption \ref{as:cost} (a). We obtain a contradiction. Hence, $\costdiff(\strategydiseq{}, \hotcap)> 0$.

Next, we prove that $\strategydiseq{\actionp} > 0$. Consider agents whose value of time satisfies $\vot \in [\votmax / 2, \votmax]$ and carpool disutility satisfies $\disu \in \left[0, \frac{1}{3}\min\left\{\votmax \costdiff(\strategydiseq{}, \hotcap), \toll\right\} \right]$. For those agents, we have $\disu \leq \frac{1}{3}\min\left\{\votmax \costdiff(\strategydiseq{}, \hotcap), \toll\right\} < \toll$, which leads to
\begin{align*}
    &C_{\actionp}(\strategydiseq{}, \vot, \disu) = \vot \costh{}(\strategydiseq{}, \hotcap) + \disu \\ 
    <& \vot \costh{}(\strategydiseq{}, \hotcap) + \toll = C_{\actiont}(\strategydiseq{}, \vot, \disu).
\end{align*}
Additionally, we have 
\begin{align*}
    C_{\actionp}(\strategydiseq{}, \vot, \disu) =& \vot \costh{}(\strategydiseq{}, \hotcap) + \disu \\
    \leq& \vot \costh{}(\strategydiseq{}, \hotcap) + \frac{1}{3} \votmax \costdiff(\strategydiseq{}, \hotcap)\\
    <& \vot \costh{}(\strategydiseq{}, \hotcap) + \vot \costdiff(\strategydiseq{}, \hotcap)\\
    =& \vot \costo(\strategydiseq{}, 1 - \hotcap) = C_{\actiono}(\strategydiseq{}, \vot, \disu),
\end{align*}
where the last inequality is due to $\vot \geq \frac{1}{2} \votmax > \frac{1}{3} \votmax$. Hence, for those agents, we have $C_{\actionp}(\strategydiseq{}, \vot, \disu) < C_{\actiono}(\strategydiseq{}, \vot, \disu)$ and $C_{\actionp}(\strategydiseq{}, \vot, \disu) < C_{\actiont}(\strategydiseq{}, \vot, \disu)$. That is, all agents with preference parameters in this set choose to carpool. Therefore, \[
    \strategydiseq{\actionp} \geq \int_{\votmax / 2}^{\votmax} \int_{0}^{\frac{1}{3}\min\left\{\votmax \costdiff(\strategydiseq{}, \hotcap), \toll\right\}} \pdf(\vot, \disu) d\disu d\vot > 0.
\]

Finally, we are going to argue that $\strategydiseq{\actiono} > 0$. Assume for the sake of contradiction that $\strategydiseq{\actiono} = 0$, then \[
    \costdiff(\strategydiseq{}, \hotcap) = \costo{}(0, 1 - \hotcap) - \costh{}(\strategydiseq{}, \hotcap) < 0,
\] which is a contradiction with the fact that $\costdiff(\strategydiseq{}, \hotcap) > 0$.
\hfill $\square$
\color{black}
\vspace{0.3cm}

\noindent\textit{Proof of Lemma \ref{lem:BestResponse}}
Recall from \eqref{eq:cost_functions}, agents whose best response is $\actiont$ satisfy $C_{\actiont}(\strategydis{}, \vot, \disu) \leq C_{\actionp}(\strategydis{}, \vot, \disu)$ and $C_{\actiont}(\strategydis{}, \vot, \disu) \leq C_{\actiono}(\strategydis{}, \vot, \disu)$, i.e. the cost of paying the toll is smaller than or equal to the cost of any other two actions. From $C_{\actiont}(\strategydis{}, \vot, \disu) \leq C_{\actionp}(\strategydis{}, \vot, \disu)$, we obtain $\vot \costh{}(\strategydis{}, \hotcap) + \toll \leq \vot \costh{}(\strategydis{}, \hotcap) + \disu$, which yields $\toll \leq \disu$. Moreover, from $C_{\actiont}(\strategydis{}, \vot, \disu) \leq C_{\actiono}(\strategydis{}, \vot, \disu)$, we obtain $\vot \costh{}(\strategydis{}, \hotcap) + \toll \leq \vot \costo{}(\strategydis{}, 1 - \hotcap)$, which yields $\toll \leq \vot \costdiff(\sigma, \hotcap)$. Thus, we obtain $\bestregion{\actiont}(\strategydis{})$. 

Similarly, agents whose best response is $\actionp$ satisfy $C_{\actionp}(\strategydis{}, \vot, \disu) = \vot \costh{}(\strategydis{}, \hotcap) + \disu \leq C_{\actiont}(\strategydis{}, \vot, \disu) = \costh{}(\strategydis{}, \hotcap) + \toll$ and $C_{\actionp}(\strategydis{}, \vot, \disu) = \vot \costh{}(\strategydis{}, \hotcap) + \disu \leq C_{\actiono}(\strategydis{}, \vot, \disu) = \vot \costo{}(\strategydis{}, 1 - \hotcap)$. It yields $\disu \leq \toll$ and $\disu \leq \vot \costdiff(\strategydis{}, \hotcap)$, and thus we obtain $\bestregion{\actionp}(\strategydis{})$. Lastly, agents whose best response is $\actiono$ satisfy $C_{\actiono}(\strategydis{}, \vot, \disu) = \vot \costo{}(\strategydis{}, 1 - \hotcap) \leq C_{\actiont}(\strategydis{}, \vot, \disu) = \vot \costh{}(\strategydis{}, \hotcap) + \toll$ and $C_{\actiono}(\strategydis{}, \vot, \disu) = \vot \costo{}(\strategydis{}, 1 - \hotcap) \leq C_{\actionp}(\strategydis{}, \vot, \disu) = \vot \costh{}(\strategydis{}, \hotcap) + \disu$. It yields $\vot \costdiff(\strategydis{}, \hotcap) \leq \min\{\toll, \disu\}$, and thus we obtain $\bestregion{\actiono}(\strategydis{})$.
\hfill $\square$
\vspace{0.3cm}

\noindent\textit{Proof of Theorem \ref{thm:NE}:} 
    In each regime, we first show that under the regime condition, the equations described in the theorem has a unique fixed point. Then, we verify that $(\sigma^*_{\actiont}, \sigma^*_{\actionp}, \sigma^*_{\actiono})$ provided for each regime indeed satisfies the equilibrium definition. 

    \medskip 
    \noindent \underline{Regime $A$}:     
    \begin{enumerate}
        \item[(A-1)] $\votmax \lthreshold \leq \ugamma$. 

        Consider a function 
        \[
        q(z) := \frac{z}{\int_{0}^{\ubeta} \int_0^{\costdiff(0, z, 1 - z, \hotcap) \beta} \pdf(\beta, \gamma) \, d\gamma d\beta}.
        \]

        Recall that under regime $A$, we have $\toll \geq \min\{\ubeta \lthreshold, \ugamma\}$. Therefore, we have $\ubeta \lthreshold \leq \min\{\toll, \ugamma\}$. Thus, for any $\beta > 0$, we have $\frac{\min\{\tau, \ugamma\}}{\ubeta}\beta \geq \lthreshold \beta = \costdiff(0, \poolmax , 1 -  \poolmax) \beta$. Therefore, 
        \begin{align*}
            &\poolmax = \int_{0}^{\ubeta} \int_0^{\frac{\min\{\tau, \ugamma\}}{\ubeta}\beta} \pdf(\beta, \gamma) \, d\gamma d\beta \\
            \geq& \int_{0}^{\ubeta} \int_0^{\costdiff(0, \poolmax , 1 -  \poolmax) \beta} \pdf(\beta, \gamma) \, d\gamma d\beta. 
        \end{align*}
        Hence, $q(\poolmax) \geq 1$. Additionally, it is easy to see that $q(0) = 0$.

        Since $\costdiff(0, z, 1 - z, \hotcap)$ is monotonically decreasing with $z$, $q(z)$ is continuous and monotonically increasing in $z$. Therefore there must be a unique point $\sigma^*_{\actionp} \in [0, \poolmax]$ such that $q(\sigma^*_{\actionp}) = 1$, which is the unique solution of equation \eqref{eq:a1-fixed_point}: 
        \[\sigma^*_{\actionp} = \int_{0}^{\ubeta} \int_0^{\costdiff(0, \sigma^*_{\actionp}, 1 - \sigma^*_{\actionp}) \beta} \pdf(\beta, \gamma) \, d\gamma d\beta.\]

        Now we want to show that $\sigma^*_{\actiont} = 0, \sigma^*_{\actionp}, \sigma^*_{\actiono} = 1 - \sigma^*_{\actionp}$ satisfies the equilibrium condition \eqref{eq: equilibrium condition}.

        First, we argue that at equilibrium, $\sigma^*_{\actiont} = 0$. From previous argument, we know that $\sigma^*_{\actionp} \leq \poolmax$. Plugging in the definitions of strategy distribution and $\poolmax$, we obtain $\int_{0}^{\ubeta} \int_0^{\costdiff(0, \sigma^*_{\actionp}, 1 - \sigma^*_{\actionp}) \beta} \pdf(\beta, \gamma) \, d\gamma d\beta \leq \int_{0}^{\ubeta} \int_0^{\frac{\min\{\tau, \ugamma\}}{\ubeta}\beta} \pdf(\beta, \gamma) \, d\gamma d\beta$, which implies $\costdiff(0, \sigma^*_{\actionp}, 1 - \sigma^*_{\actionp}) \leq \frac{\min\{\tau, \ugamma\}}{\ubeta}$. Therefore, for any $(\vot, \disu)$, we have $\vot \costdiff(0, \sigma^*_{\actionp}, 1 - \sigma^*_{\actionp}) \leq \min\{\toll, \ugamma\}$. Hence, for all agents, we have $C_{\actiono}(\strategydis{}, \vot, \disu) \leq C_{\actiont}(\strategydis{}, \vot, \disu)$. Thus, no agents want to deviate to pay the toll and we have $\sigma^*_{\actiont} = 0$ at equilibrium.
        

        Next, we argue that at equilibrium, the strategy distribution of carpooling agents is given by $\sigma^*_{\actionp}$. An agent $(\beta,\gamma)$ will choose $\actionp$ over $\actiono$ if $C_{\actionp}(\strategydiseq{}, \vot, \disu) < C_{\actiono}(\strategydiseq{}, \vot, \disu)$, which is equivalent to $\disu < \vot \costdiff(0, \sigma^*_{\actionp}, 1 - \sigma^*_{\actionp})$. By lemma \ref{lem:BestResponse}, the set of agents whose best response is to carpool under regime A-1 is then given by $\bestregion{\actionp}(\strategydiseq{}) = \{(\beta, \gamma) : 0 < \beta < \ubeta, 0 < \gamma < \beta \costdiff(0, \sigma^*_{\actionp}, 1 - \sigma^*_{\actionp})\}$. Integrating over all agents in $\bestregion{\actionp}(\strategydiseq{})$ yields the $\sigma^*_{\actionp}$ given above. Lastly, because we have argued that $\sigma^*_{\actiont} = 0$ and $\sigma^*_{\actionp}$ satisfy the equilibrium condition \eqref{eq: equilibrium condition}, we can obtain that $\sigma^*_{\actiono} = 1 - \sigma^*_{\actionp}$ must also satisfy the equilibrium condition. Hence, we have proved that $(\sigma^*_{\actiont}, \sigma^*_{\actionp}, \sigma^*_{\actiono})$ provided for regime A-1 satisfies the equilibrium condition.
        

        \item[(A-2)] $\ubeta \lthreshold > \ugamma$.

        Consider the following function $g$:
        \[
        	g(z) := \frac{1 - z}{\int_0^{\ugamma} \int_{0}^{\frac{\gamma}{\costdiff(0, z, 1 - z, \hotcap)}} \pdf(\beta, \gamma) d\beta \, d\gamma}.
        \]

        We first argue that $g(\poolmax) > 1$. From the regime A condition such that $\tau \geq \min\left\{\ugamma, \votmax \lthreshold \right\}$, we obtain $\votmax \lthreshold > \min\{\toll, \ugamma\}$ and $\toll \geq \ugamma$.         Then, we have
        \begin{align*}
            1 - \poolmax =& 1 - \int_{0}^{\ubeta} \int_0^{\frac{\min\{\tau, \ugamma\}}{\ubeta}\beta} \pdf(\beta, \gamma) \, d\gamma d\beta\\
            \stackrel{(a)}{>}& 1 - \int_{0}^{\ubeta} \int_0^{\beta \costdiff(0, \poolmax , 1 -  \poolmax)} \pdf(\beta, \gamma) \, d\gamma d\beta\\
            =& \int_{0}^{\ubeta} \int_{\beta \costdiff(0, \poolmax, 1 - \poolmax)}^{\ugamma} \pdf(\beta, \gamma) \, d\gamma d\beta\\
            \stackrel{(b)}{=}& \int_0^{\ugamma} \int_{0}^{\frac{\gamma}{\costdiff(0, \poolmax, 1 - \poolmax, \hotcap)}} \pdf(\beta, \gamma) d\beta \, d\gamma
        \end{align*}
        (a) is due to the inequality $\min\{\toll, \ugamma\} < \votmax \lthreshold = \votmax \costdiff(0, \poolmax , 1 -  \poolmax)$, and (b) is obtained by changing the integration order. Therefore, we obtain $g(\poolmax) > 1$. Additionally, it is easy to see that $g(1) = 0$. Since $\costdiff(0, z, 1 - z, \hotcap)$ is monotonically decreasing with $z$, we know that $g(z)$ is monotonically decreasing with $z$. Therefore, there exist a unique $\sigma^*_{\actionp}> \poolmax$ such that $g(\sigma^*_{\actionp}) = 1$, which means that it is the unique solution of equation \eqref{eq:a2-fixed_point}: 
        \[1- \sigma^*_{\actionp} = \int_0^{\ugamma} \int_{0}^{\frac{\gamma}{\costdiff(0, \sigma^*_{\actionp}, 1-\sigma^*_{\actionp})}} \pdf(\beta, \gamma) d\beta d\gamma. \]

        Next, we are going to show that $\sigma^*_{\actiont} = 0, \sigma^*_{\actionp}, \sigma^*_{\actiono} = 1 - \sigma^*_{\actionp}$ satisfies the equilibrium condition \eqref{eq: equilibrium condition}.

        Due to the condition of regime A-2 such that $\toll \geq \ugamma$, we obtain that $C_{\actionp}(\strategydis{}, \vot, \disu) \leq C_{\actiont}(\strategydis{}, \vot, \disu)$ holds for all agents. Therefore, no agents want to deviate to toll paying and thus $\sigma^*_{\actiont} = 0$ at equilibrium. Additionally, by lemma \ref{lem:BestResponse}, the set of agents using ordinary lanes under regime A-2 is given by $\bestregion{\actiono}(\strategydiseq{}) = \{(\beta, \gamma): \beta  \leq \gamma/\costdiff(\sigma^*)\}$. Hence, under equilibrium, we obtain 
        \begin{align*}
            &\sigma^*_{\actiono} = \iint_{\Lambda_{\actiono}(\sigma^*)} \pdf(\beta,\gamma) d\beta d\gamma \\
            =& \int_0^{\ugamma} \int_{0}^{\frac{\gamma}{\costdiff(0, \sigma^*_{\actionp}, 1-\sigma^*_{\actionp})}} \pdf(\beta, \gamma) d\beta d\gamma = 1 - \sigma^*_{\actionp},
        \end{align*}
        which is equivalent to equation \eqref{eq:a1-fixed_point}. Therefore, $\sigma^*_{\actionp}$ and $\sigma^*_{\actiono} = 1 - \sigma^*_{\actionp}$ satisfy the equilibrium condition \ref{eq: equilibrium condition} as well.
    \end{enumerate}

    \noindent \underline{Regime $B$}: 

    To show that the system of equations \eqref{eq:b-fixed_point} has a unique solution, we first represent the strategy distributions using the latency difference between ordinary lanes and HOT lanes, which we denote as $\delta$. Consider function $g: [\toll/\votmax, \infty) \rightarrow [0, \bar{\sigma}_{\actiont})$, where $\bar{\sigma}_{\actiont}$ is defined as $\bar{\sigma}_{\actiont} := \int_{\tau}^{\ugamma} \int_{0}^{\ubeta} \pdf(\beta, \gamma) \, d\beta  d\gamma$, \[
        g(y) = \int_{\tau}^{\ugamma} \int_{\toll/y}^{\ubeta} \pdf(\beta, \gamma) \, d\beta  d\gamma,
    \] and function $h$ with the domain $[\toll/\votmax, \infty)$: \[
        h(y) = \int_0^{\tau} \int_{\gamma / y}^{\votmax} \pdf(\beta, \gamma) \, d\beta d\gamma.
    \] We note that both $g$ and $h$ are continuous and increase monotonically with $y$. Moreover, we note that the range of $h$ is $[\poolmax, 1 - \bar{\sigma}_{\actiont})$, because 
    \begin{align*}
        h(\tau/\votmax) =& \int_{0}^{\tau} \int_{\ubeta \gamma / \tau}^{\votmax} f(\beta, \gamma) d\beta d\gamma \\
        =& \int_{0}^{\votmax} \int_{0}^{\beta \tau/\votmax} f(\beta, \gamma) d\gamma d\beta\\
        =& \int_{0}^{\votmax} \int_{0}^{\min\{\tau, \ugamma \} \beta/\votmax} f(\beta, \gamma) d\gamma d\beta = \poolmax,
    \end{align*} and 
    \begin{align*}
        &\lim_{\delta \rightarrow \infty} h(\delta) = \int_{0}^{\tau} \int_{0}^{\votmax} f(\beta, \gamma) d\beta d\gamma \\
        =& 1 - \int_{\tau}^{\ugamma} \int_{0}^{\votmax} f(\beta, \gamma) d\beta d\gamma = 1 - \bar{\sigma}_{\actiont}.
    \end{align*}
    
    Let $\delta^*$ be the induced latency difference at equilibrium. By equations \eqref{eq:b-fixed_point}, we obtain that $\sigma^*_{\actiont} = g(\delta^*)$, $\sigma^*_{\actionp} = h(\delta^*)$, and $\sigma^*_{\actiono} = 1 - g(\delta^*) - h(\delta^*)$. Hence, in order to show that \eqref{eq:b-fixed_point} has a unique solution, it is suffice to show that there exists a solution to the following fixed point equation:
    \begin{equation} \label{eq:b-fixed_point_reduced-delta}
        \delta = \costdiff(g(\delta), h(\delta), 1 - g(\delta) - h(\delta)).
    \end{equation}
    
    Define the function $q$ with the domain $[\tau/\votmax, \infty)$: \[
        q(\delta) = \frac{\costdiff(g(\delta), h(\delta), 1 - g(\delta) - h(\delta))}{\delta}.
    \] We remark that $q$ is continuous and mononotically decreasing in $\delta$. First, we have 
    \begin{align*}
        q(\tau/\votmax) =& \frac{\costdiff(g(\tau/\votmax), h(\tau/\votmax), 1 - g(\tau/\votmax) - h(\tau/\votmax))}{\tau/\votmax}\\ 
        \stackrel{(a)}{=}& \frac{\votmax \lthreshold}{\tau} \stackrel{(b)}{>} 1,
    \end{align*}
    where (a) is due to $g(\tau/\votmax) = 0$ and $h(\tau/\votmax) = \poolmax$;
    (b) is due to the regime condition $\tau < \votmax \lthreshold$. Furthermore, it is easy to see that as $\delta \uparrow \infty$, $q(\delta) \rightarrow 0$. Since $q$ is continuous monotonically decreasing, there exists a unique $\delta^* \in [\tau/\votmax, \infty)$ such that $q(\delta^*) = 1$, which implies that there exists a unique solution to \eqref{eq:b-fixed_point_reduced-delta}. Hence, there exists a unique solution to \eqref{eq:b-fixed_point}.

    Given the strategy distribution $\sigma^*$, integrating over the best response region for each of the three actions given in lemma \ref{lem:BestResponse} yields exactly the system of equations \eqref{eq:b-fixed_point}. Therefore, the solution obtained in $\eqref{eq:b-fixed_point}$ is an equilibrium.
\hfill $\square$
\vspace{0.3cm}

\noindent\textit{Proof of Theorem \ref{theorem:comparative}:} \label{sec:proof-comparative}

\noindent \underline{Fix $\toll$ and increase $\hotcap$}: 

    We first show that $\costdiff(\strategydiseq{}, \hotcap)$ increases. We assume for the sake of contradiction that $\costdiff(\strategydiseq{}, \hotcap)$ decreases. Then, from lemma \ref{lem:BestResponse}, more agents will join the ordinary lanes, and fewer  will carpool or pay the toll price. That is, $\sigma^*_{\actiono}$ increases, $\sigma^*_{\actionp}$ decreases, and $\sigma^*_{\actiont}$ is non-increasing. Recall from assumption \ref{as:cost} that the latency function for ordinary lanes (resp. HOT lanes) increases (resp. decreases) with $\hotcap$. Therefore, $\costo{}(\strategydiseq{\actiono} \demand{}{}, 1 - \hotcap)$ increases as both $\strategydiseq{\actiono}$ and $\hotcap$ increase. Moreover, $\costh{}\left( \left(\strategydiseq{\actiont} + \frac{\strategydiseq{\actionp}}{\poolsize}\right) D, \hotcap\right)$ decreases because the term $\left(\strategydiseq{\actiont} + \frac{\strategydiseq{\actionp}}{\poolsize}\right)$ decreases and $\hotcap$ increases. Hence, $\costdiff(\strategydiseq{}, \hotcap)$ increases, which is a contradiction. 
    
    Next, we assume for the sake of contradiction that $\costdiff(\strategydiseq{}, \hotcap)$ does not change. Using lemma \ref{lem:BestResponse}, we obtain that the strategy distributions for all three actions should hold fixed. Nevertheless, given the same strategy distributions, increasing $\hotcap$ leads to an increase in $\costo{}(\strategydiseq{\actiono} \demand{}{}, 1 - \hotcap)$ and a decrease in $\costh{}\left( \left(\strategydiseq{\actiont} + \frac{\strategydiseq{\actionp}}{\poolsize}\right) D, \hotcap\right)$. This yields an increase in $\costdiff(\strategydiseq{}, \hotcap)$, which is a contradiction.
    


    Since $\costdiff(\strategydiseq{}, \hotcap)$ increases, from lemma \ref{lem:BestResponse}, we obtain that the strategy distributions $\sigma^*_{\actiono}$ decreases, $\sigma^*_{\actionp}$ increases, and $\sigma^*_{\actiont}$ increases. 

\noindent \underline{Fix $\hotcap$ and increase $\toll$}: 

    We first show that at equilibrium $\costdiff(\strategydiseq{}, \hotcap)$ is non-decreasing. Assume for the sake of contradiction that $\costdiff(\strategydiseq{}, \hotcap)$ decreases. Since $\toll$ increases, from lemma \ref{lem:BestResponse}, we obtain that the population paying toll price decreases, the population taking the ordinary lanes is increases, and the population that carpools decreases. Therefore, $\costdiff(\strategydiseq{}, \hotcap)$ increases, which is a contradiction.

    Last but not least, we argue that $\strategydiseq{\actiono}$ can change in either direction when $\toll$ increases. When the toll price is relatively large, i.e. in regime A in theorem \ref{thm:NE}, further increasing the $\toll$ has no impact on the strategy distributions, and hence $\strategydiseq{\actiono}$ does not chance. When the toll price is relatively low, i.e. in regime B in theorem \ref{thm:NE}, an increase in $\toll$ can either lead to an increase or decrease of the term $\frac{\toll}{\costdiff(\strategydiseq{}, \hotcap)}$ in \eqref{eq:b-fixed_point-toll}, which depends on the preference distribution. Therefore, $\strategydiseq{\actiono}$ can also change in either direction given the increase in $\toll$.
\hfill $\square$

\medskip

\noindent \textit{Proof of Theorem \ref{thm:multiseg-equi-uniqueness}.}
We note that for each segment $\segnum \in [\segtotal]$, $\delta_e \in [\underline{\delta}_e, \bar{\delta}_e]$, where the lower bound $\underline{\delta}_e= \ell_{o, e}(0) -\costh{,\segnum}\left(\sum_{i = 1}^{\segnum} \sum_{j = \segnum}^{\segtotal} D^{ij} \right)$ is achieved when all agents choose the HOT lane without pooling, and the upper bound $\bar{\delta}_e=\costo{, \segnum}\left(\sum_{i = 1}^{\segnum} \sum_{j = \segnum}^{\segtotal} D^{ij} \right)- \ell_{e, h}(0)$ is achieved when all agents take the ordinary lane. Moreover, since the preference distribution function $f(\beta, \gamma)$ is continuous, the function $\Phi: \prod_{e \in E} [\underline{\delta}_e, \bar{\delta}_e] \to \prod_{e \in E} [\underline{\delta}_e, \bar{\delta}_e]$ is continuous. From the Kakutani's fixed-point theorem, we know that the set of fixed-points of $\Phi(\delta)=\delta$ is non-empty. Furthermore, for any such fixed point $\delta^*$, the associated best-response strategy $s^*$ is unique, and satisfies the condition in Definition \ref{def:multiseg-eq}, and thus is a Wardrop equilibrium. On the other hand, for any Nash equilibrium $s^*$, the induced equilibrium latency cost difference must satisfy $\Phi(\delta^*) = \delta^*$. Therefore, the fixed-point set of $\Phi(\delta)=\delta$ and the set of Wardrop equilibrium has one-to-one correspondence. Since the fixed-point set is non-empty, we can conclude that Wardrop equilibrium exists. 

Additionally, equilibrium is unique if and only if $\Phi(\delta)=\delta$ has a unique solution. If there exist two fixed points $\delta, \delta'$ satisfying $\Phi(\delta) = \delta$ and $\Phi(\delta') = \delta'$. Let $v_i$ denote the $|\segtotal|$-dimensional vector with $i$-th index being $1$ and all other indices being $0$. Then, by the mean value theorem, there exists $z$ on the line segment connecting $\delta$ and $\delta'$ such that \[
    v_e \cdot [\triangledown \Phi(z) (\delta - \delta')] = v_e \cdot [\Phi(\delta) - \Phi(\delta')], \quad \forall e \in E.
\] Combining the equations for all $e \in \segtotal$, we obtain \begin{align}\label{eq:eigenvalue}
    \triangledown \Phi(z) (\delta - \delta') = \Phi(\delta) - \Phi(\delta') = \delta - \delta'.
\end{align}
If the Jacobian $\nabla \Phi(z)$ does not have eigenvalue 1 for any $z$, then there do not exist $\delta, \delta'$ that satisfies \eqref{eq:eigenvalue}, and thus equilibrium is unique. \hfill $\square$